\documentclass[american,nofootinbib]{revtex4-2}
\usepackage{lmodern}

\usepackage[T1]{fontenc}
\usepackage{geometry}
\usepackage{color}
\usepackage{babel}
\usepackage{array}
\usepackage{enumitem}
\usepackage{multirow}
\usepackage{amsmath}
\usepackage{amsthm}
\usepackage{amssymb}
\usepackage{graphicx}
\usepackage{wasysym}
\usepackage{dsfont}
\usepackage[normalem]{ulem}
\usepackage[unicode=true,pdfusetitle,
 bookmarks=true,bookmarksnumbered=false,bookmarksopen=false,
 breaklinks=false,pdfborder={0 0 0},pdfborderstyle={},backref=false,colorlinks=false]
 {hyperref}

\makeatletter

\providecommand{\tabularnewline}{\\}

\@ifundefined{showcaptionsetup}{}{%
 \PassOptionsToPackage{caption=false}{subfig}}
\usepackage{subfig}
\makeatother

\makeatother

\begin{document}
\title{ Non-comoving description of adiabatic radial perturbations of relativistic stars}
\author{Paulo Luz}
\email{paulo.luz@tecnico.ulisboa.pt}

\affiliation{Centro de Astrof\'{\i}sica e Gravita\c{c}\~{a}o - CENTRA, Departamento
de F\'{\i}sica, Instituto Superior T\'{e}cnico - IST, Universidade
de Lisboa - UL, Av. Rovisco Pais 1, 1049-001 Lisboa, Portugal,}
\affiliation{Departamento de Matem\'{a}tica, ISCTE - Instituto Universit\'{a}rio
de Lisboa, Portugal}
\author{Sante Carloni}
\email{sante.carloni@unige.it}

\affiliation{Institute of Theoretical Physics, Faculty of Mathematics and Physics,
Charles University, Prague, V Hole\v{s}ovi\v{c}k\'{a}ch 2, 180 00
Prague 8, Czech Republic,}
\affiliation{DIME, Universit\`{a} di Genova, Via all'Opera Pia 15, 16145 Genova,
Italy,}
\affiliation{INFN Sezione di Genova, Via Dodecaneso 33, 16146 Genova, Italy}
\begin{abstract}
We study adiabatic, radial perturbations of static, self-gravitating
perfect fluids within the theory of general relativity employing a
new perturbative formalism. We show that by considering a radially
static observer, the description of the perturbations can be greatly
simplified with respect to the standard comoving treatment. 
The new perturbation equations can be solved to derive
analytic solutions to the problem for a general class
of equilibrium solutions. We discuss the thermodynamic description
of the fluid under isotropic frame transformations, showing how, in
the radially static, non-inertial frame, the stress-energy tensor
of the fluid must contain momentum transfer terms. As illustrative
examples of the new approach, we study perturbations of equilibrium
spacetimes characterized by the Buchdahl I, Heintzmann IIa, Patwardhan-Vaidya
IIa, and Tolman VII solutions, computing the first oscillation eigenfrequencies
and the associated eigenfunctions. We also analyze
the properties of the perturbations of cold neutron stars composed
of a perfect fluid verifying the Bethe-Johnson model I equation of
state, computing the oscillation eigenfrequencies and the $e$-folding
time.
\end{abstract}
\maketitle

\section{Introduction}
Perturbative analysis, when possible, is one of the most important tools for exploring the intricacies of complex physical phenomena. In general relativity (GR), the study of perturbative solutions is crucial in several different contexts. Nonetheless, a fully mature theory of perturbation for relativistic gravitation came quite late compared to other research sectors. This delay can be ascribed to two specific issues in developing the perturbation theory: the so-called gauge problem and the consistent formulation of relativistic thermodynamics. Indeed, the fundamental dependence of a relativistic perturbation theory on the gauge makes it challenging to understand the behavior of the perturbed spacetime, possibly leading to ambiguous conclusions~\cite{Stewart_Walker_1974}.

In the study of black holes perturbations, the first gauge-invariant approach was proposed by Moncrief~\cite{Montcrief_1974} and many works have further developed the subject, in particular, Ref.~\cite{Clarkson_Barrett_2003} developed a gauge-invariant and covariant framework to study linear perturbations of static, spherically symmetric black holes. In the context of cosmology, Bardeen pioneered the development of a gauge-independent approach to perturbations~\cite{Bardeen:1980kt}, and Bruni and Ellis have constructed a fully covariant, gauge-invariant theory of cosmological perturbations~\cite{Ellis:1989jt}. In relativistic astrophysics, however, until recently, there was no covariant, gauge-invariant perturbation theory applicable to those types of solutions.

In the case of relativistic compact stellar objects, all studies on the subject rely on one specific gauge. Moreover, in many instances, the gauge is chosen to itself encode completely the degrees of freedom of the perturbations. This approach was first introduced by Chandrasekhar while studying adiabatic, radial perturbations of stellar compact objects~\cite{Chandrasekhar_1964_PRL, Chandrasekhar_1964_ApJ}, and all subsequent studies and reformulations have adopted this idea~\cite{Chanmugam_1977, Gondek_1997, Kokkotas_Ruoff_2001}. Although undeniably pioneering, the Chandrasekhar approach presents many issues and limitations. One such limitation is the focusing of the description of adiabatic, radial perturbations to the point of view of the frame locally comoving with the fluid, hindering the possibility of exploiting covariance to simplify the perturbation problem.

Given the revolution that multimessenger astronomy has brought to relativistic astrophysics, it becomes increasingly important to develop perturbation schemes that can perform better than the ones currently in use. In Ref.~\cite{Luz_Carloni_2024a}, we have constructed such a scheme by developing a full covariant, gauge-invariant theory of perturbation for nonvacuum, compact, locally rotationally symmetric of class II spacetimes, which can be used to study perturbations of relativistic stars spacetimes. Specifically, we laid out the mathematical foundations of the perturbation theory, addressing the properties of the differential equations that describe general perturbations, and proposed a method to find exact solutions in terms of power series. A second paper, Ref.~\cite{Luz_Carloni_2024b}, was dedicated to the application of the new perturbation theory considering the locally comoving frame with the perturbed fluid. We studied perturbations of classical exact solutions of the Einstein field equations, computing the first eigenfrequencies and deducing universal criteria for stability. The analysis, however, confirmed what was already discussed in Ref.~\cite{Luz_Carloni_2024a}: that formulation of the problem is not the most computationally efficient, whereas a change to a non-comoving frame would be more advantageous.

The aim of the present paper is then to explore the potential of the static frame perturbation equations. A challenge in this new formulation will be the identification of the thermodynamic properties of the matter sources in a static, non-inertial frame. However, the covariance of the formalism will allow us to characterize the effects, offering, at the same time, 
several interesting insights into relativistic thermodynamics.

The results found in this paper, as well as Refs.~\cite{Luz_Carloni_2024a} and \cite{Luz_Carloni_2024b} rely on a specific formalism called 1+1+2 covariant approach developed in Ref.~\cite{Clarkson_Barrett_2003,Betschart_Clarkson_2004,Clarkson_2007}. The 1+1+2 covariant approach can be thought of as an extension of the so-called 1+3 decomposition (cf., e.g., Refs.~\cite{Ellis:1998ct, Luz_Lemos_2023}). Both these formalisms can be considered semi-tetradic representations of the spacetime, where all tensor quantities are locally, partially decomposed with respect to a timelike and a spacelike vector field. The advantage of this description of gravitational systems is manifold, but probably the most important is the ability to describe in a relatively simple, mathematically rigorous, and physically transparent very complex
gravitational systems without relying on specific coordinate systems.
The 1+1+2 covariant approach, in particular, has been used to formulate a covariant version of the Tolman-Oppenheimer-Volkoff (TOV) equations, leading to the discovery of new solutions in GR
~\cite{Carloni_Vernieri_2018a,Carloni_Vernieri_2018b,Naidu:2021nwh,Naidu:2022igk}, and in some extension of Einstein's theory~\cite{Luz_Carloni_2019,Campbell:2024ouk}, and used to reformulate the theory of junction conditions~\cite{Rosa:2023tbh}.

The paper is organized as follows. Section~\ref{sec:Thermodynamic-description}  describes in detail the thermodynamics of a perfect fluid in a non-comoving frame, emphasizing the first law of thermodynamics, heat transfer, and entropy transformation. Section~\ref{sec:AdiabaticRadialPerturbations} contains a description of the perturbation variables, the perturbation equations together with their boundary condition, and a general algorithm to find power series solutions of the perturbation equations. Section~\ref{sec:Examples} deals with the application of the resolution algorithm to some specific cases that are not easily solvable in the comoving frame. In Section~\ref{sec:Perturbation_of_neutron_stars}, we analyze the properties of the adiabatic, radial perturbations of cold neutron stars composed by a fluid characterized by the Bethe-Johnson model I equation of state. We then draw our conclusions in
Section~\ref{sec:Conclusions}. The paper contains two appendices. In Appendix~\ref{Appendix:sec_1p1p2_decomposition},
we introduce the general definitions for the 1+1+2 covariant quantities, and in Appendix~\ref{Appendix:sec_isotropic_frame_transformations}, we present the general transformation associated with a change of frame and its consequences on the 1+1+2 quantities.

Throughout the article we will consider the metric signature $\left(-+++\right)$, and, except in Section~\ref{sec:Perturbation_of_neutron_stars}, we will work in the geometrized unit system
where $8\pi G=c=k_{B}=1$.

\section{\label{sec:Thermodynamic-description}Thermodynamic description of a perfect fluid in the static frame}

We are interested in studying perturbative solutions of static, spatially
compact, spherically symmetric spacetimes with a perfect fluid source
under adiabatic, radial perturbations using the new formalism constructed
in Ref.~\cite{Luz_Carloni_2024a}. In Ref.~\cite{Luz_Carloni_2024b}
this analysis was carried out considering the point of view of a comoving
observer, the so-called {\it Lagrangian picture} of the fluid flow. In this
article, we will show how the analysis can be greatly simplified if,
instead,  we consider the point of view of a static observer with respect to a static observer at spatial infinity,
which can be considered as the gauge-invariant, covariant {\it Eulerian picture}.
In that regard, we have to allow a
general stress-energy tensor for the perturbed fluid. We will impose
the perturbed fluid to remain perfect. However, as it is shown in Appendix~\ref{Appendix:sec_isotropic_frame_transformations}, from the point of view of a static observer, the fluid
will not be seen as perfect, and momentum transfer terms have to be
included. However, in this section, we will show that in first-order perturbation theory, those momentum
fluxes do not represent energy exchange between the volume elements
of the fluid as heat. This discussion will help the understanding of the results on the evolution of the perturbations contained in the next sections.

To keep the treatment in this section independent of the choice of
a local coordinate system, we will consider the covariant quantities
of the 1+1+2 formalism, whose definitions we present in Appendix~\ref{Appendix:sec_1p1p2_decomposition}.
Nonetheless, the final set of perturbation equations in the static
frame used in the following sections will be written immediately in
the standard Schwarzschild coordinate system and, thus, no knowledge of covariant
methods is required for their use.

\subsection{Transformation of the particle number density vector field}

Consider a perfect fluid permeating a spacetime and two observers:
an observer locally comoving with the volume elements of the fluid
and another observer whose frame is related with the comoving frame
by an isotropic frame transformation (cf. Appendix~\ref{Appendix:sec_isotropic_frame_transformations}).
Let $u$ represent the 4-velocity of the comoving observer and $e$
represent a spacelike vector field orthogonal to $u$, such that the
integral curves of $u$ and the integral curves of $e$ form two local
congruences in the spacetime. On the other hand, let an observer in
the second frame be characterized by a 4-velocity $\bar{u}$ and $\bar{e}$
be a spacelike vector field orthogonal to $\bar{u}$, such that, similarly,
the integral curves of $\bar{u}$ and the integral curves of $\bar{e}$
also form two local congruences in the spacetime. In short, we will
say the dyad $\left(u,e\right)$ is associated with the comoving frame
and $\left(\bar{u},\bar{e}\right)$ is associated with the second
frame. In what follows, for simplicity, we will use an overline to
identify quantities defined in the second frame. In addition, without
loss of generality, we will assume the integral curves of the vector
fields to be affinely parameterized, and the tangent vector fields are normalized
such that
\begin{equation}
\begin{aligned}u^{\alpha}u_{\alpha} & =-1\,, & e^{\alpha}e_{\alpha} & =1\,,\\
\bar{u}^{\alpha}\bar{u}_{\alpha} & =-1\,, & \bar{e}^{\,\alpha}\bar{e}_{\alpha} & =1\,.
\end{aligned}
\end{equation}

In the considered setup, by definition,  the particle 4-current density 
vector in the comoving frame is given by
\begin{equation}
N^{\alpha}=nu^{\alpha}\,,\label{eq:Particle_density_vector_general_definition}
\end{equation}
where $n$ represents the particle number density, whereas in the
barred frame, in general, we have
\begin{equation}
\bar{N}^{\alpha}=\bar{n}\bar{u}^{\alpha}+\bar{n}^{\alpha}\,,\label{eq:Particle_density_vector_Bar_general_definition}
\end{equation}
where $\bar{n}^{\alpha}$ is the particle number density flux vector,
orthogonal to $\bar{u}$, measured by the observer associated with
the dyad $\left(\bar{u},\bar{e}\right)$, such that $\bar{n}^{\alpha}\bar{u}_{\alpha}=0$.
Given an isotropic change of frame, Eq.~(\ref{eq:Frame_transformations_ue_noBar_to_Bar}),
the components of the vector field $N$ transform as
\begin{equation}
N^{\alpha}=nu^{\alpha}\to\bar{N}^{\alpha}=\bar{u}^{\alpha}n\cosh\beta-\bar{e}^{\,\alpha}n\sinh\beta\,,
\end{equation}
therefore,
\begin{equation}
\begin{aligned}\bar{n} & =n\cosh\beta\,,\\
\bar{n}^{\alpha} & =-\bar{e}^{\,\alpha}n\sinh\beta\,.
\end{aligned}
\label{eq:Particle_density_Bar_NoBar_relations}
\end{equation}

In what follows, we will impose particle number conservation in both
frames, that is $\nabla_{\alpha}N^{\alpha} =0$ and $\nabla_{\alpha}\bar{N}^{\alpha} =0$.
This implies the following relations for the comoving and the barred
frame, respectively,
\begin{align}
u^{\alpha}\nabla_{\alpha}n+n\theta & =0\,,\label{eq:Particle_density_conservation}\\
\bar{u}^{\alpha}\nabla_{\alpha}\bar{n}+\bar{n}\bar{\theta}+\nabla_{\alpha}\bar{n}^{\alpha} & =0\,,\label{eq:Particle_density_conservation_bar}
\end{align}
where $\theta$ and $\bar{\theta}$ represent the expansion scalars
associated with the integral curves of $u$ and of $\bar{u}$, respectively.
Using Eqs.~(\ref{eq:Frame_transformations_ue_Bar_to_noBar}), (\ref{eq:Frame_transformations_thetaBar_theta_general})
and (\ref{eq:Particle_density_conservation}), Eq.~(\ref{eq:Particle_density_conservation_bar})
can be rewritten in the useful form
\begin{equation}
\frac{\nabla_{\alpha}\bar{n}^{\alpha}}{\bar{n}}=-\sinh\beta\left[\mathcal{A}+\phi+\frac{e^{\alpha}\nabla_{\alpha}n}{n}\right]-\tanh\beta~\bar{u}^{\alpha}\nabla_{\alpha}\beta-\bar{e}^{\,\alpha}\nabla_{\alpha}\beta\,.\label{eq:Particle_density_conservation_bar_alt}
\end{equation}

\subsection{First law of thermodynamics and energy conservation in the barred
frame}

To describe the thermodynamics of the perturbed fluid, we will 
consider the local-thermodynamic equilibrium ansatz, which presupposes that	
given an off-thermodynamic equilibrium system, we can still locally define volume elements that act as a thermodynamic subsystem in equilibrium, such that within these elements, the state is characterized by well-defined state variables.
In our case, this assumption is reasonable, given the 
considered perturbative setup, and assuming that the evolution time-scale of the 
perturbations within the fluid is much greater than a characteristic local 
relaxation time.

Now, assume the fluid to be composed of a single particle species
with rest mass $m_{N}$. For the barred frame, let $\bar{\mu}$ represent
the relativistic energy density of the fluid, $\bar{\varepsilon}$
the specific internal energy density, $\bar{n}$ the particle density, and $\bar{\mu}_{N}=m_{N}\bar{n}$
the rest mass density, such that
$\bar{\mu}=\bar{\mu}_{N}\left(1+\bar{\varepsilon}\right)$,
hence
\begin{equation}
d\bar{\mu}=\left(1+\bar{\varepsilon}\right)d\bar{\mu}_{N}+\bar{\mu}_{N}d\bar{\varepsilon}\,.\label{eq:Heat_transfer_differential_relativistic_energy_static_frame}
\end{equation}
Defining the specific volume $\bar{v}=1/\bar{\mu}_{N}$, such that
\begin{equation}
d\bar{v}=-\bar{v}^{2}d\bar{\mu}_{N}=-m_{N}\bar{v}^{2}d\bar{n}\,,\label{eq:Heat_transfer_differential_v_n_relation}
\end{equation}
Eq.~(\ref{eq:Heat_transfer_differential_relativistic_energy_static_frame})
can be rewritten as
\begin{equation}
d\bar{\mu}=-\bar{\mu}\frac{d\bar{v}}{\bar{v}}+\bar{\mu}_{N}d\bar{\varepsilon}\,.
\end{equation}

For a quasi-equilibrium process, the first law of thermodynamics reads
\begin{equation}
d\bar{\varepsilon}=-\delta\overline{\mathsf{W}}+\delta\bar{\mathsf{Q}}=-\bar{p}d\bar{v}+ \overline{T\vphantom{\bar{t}}} d\bar{S}\,,
\end{equation}
where
$\delta$ represents the variation of a path-dependent quantity under a given thermodynamical process,
$\bar{p}$ represents the local pressure, $\overline{T\vphantom{\bar{t}}}$ the local temperature, and $\bar{S}$ the local specific entropy, all defined in the barred
frame. Then,
\begin{equation}
d\bar{\mu}=-\left(\bar{\mu}+\bar{p}\right)\frac{d\bar{v}}{\bar{v}}+\bar{\mu}_{N}\overline{T\vphantom{\bar{t}}} d\bar{S}\,.
\label{eq:Heat_transfer_dot_energy_density_static_frame_differential_form}
\end{equation}
From Eq.~(\ref{eq:Heat_transfer_differential_v_n_relation}) we have
\begin{equation}
\frac{d\bar{v}}{\bar{v}}=-\frac{d\bar{n}}{\bar{n}}\,.
\end{equation}
Then, along the world-lines of the barred observers and imposing
the conservation of particle density~(\ref{eq:Particle_density_conservation_bar}),
\begin{equation}
\frac{\dot{\bar{v}}}{\bar{v}}=-\frac{\dot{\bar{n}}}{\bar{n}}=\bar{\theta}+\frac{1}{\bar{n}}\nabla_{\alpha}\bar{n}^{\alpha}\,,
\end{equation}
where ``dot'' over a barred quantity represents the directional
derivative taken with respect to $\bar{u}$, that is, a derivative taken
with respect to the proper time, $\tau$, of the barred observer. Then, from Eq.~(\ref{eq:Heat_transfer_dot_energy_density_static_frame_differential_form})
we find
\begin{equation}
\dot{\bar{\mu}}=-\left(\bar{\mu}+\bar{p}\right)\left(\bar{\theta}+\frac{1}{\bar{n}}\nabla_{\alpha}\bar{n}^{\alpha}\right)+\bar{\mu}_{N}\overline{T\vphantom{\bar{t}}} \dot{\bar{S}}\,.
\label{eq:Heat_transfer_dot_energy_density_static_frame}
\end{equation}

\subsection{Perturbative analysis of the first law}

So far, the discussion was kept rather general. We will now and for
the remaining of the section consider the specific setup of an equilibrium
perfect fluid in a static, spherically symmetric spacetime that is
adiabatically, radially perturbed, such that the perturbed fluid is
also a perfect fluid. The equilibrium fluid is characterized by its
energy density, $\mu_{0}$, and pressure, $p_{0}$. Moreover, we will
assume that an observer comoving with the equilibrium fluid has
4-velocity $u_{0}$ and we can define a spacelike vector field $e_{0}$ orthogonal
to $u_{0}$ at each point. Moreover, we assume that the integral curves
of $u_{0}$ and the integral curves of $e_{0}$ form two local congruences
in the equilibrium background spacetime.

Because of the nature of the background spacetime, observers comoving with 
volume elements of the equilibrium fluid are also radially static observers.
However, in the perturbed spherically symmetric spacetime, these observers
are distinct. Let the barred frame of the previous subsections 
be associated with a radially static observer in the perturbed spacetime,
such that $\bar{e}$ is aligned with the outward radial gradient.
Then, the particle density flux vector field $\bar{n}^{\alpha}$,
defined in Eqs.~(\ref{eq:Particle_density_vector_Bar_general_definition})--(\ref{eq:Particle_density_Bar_NoBar_relations}),
vanishes in the background spacetime, since the comoving and the static
observers are the same. Therefore, $\bar{n}^{\alpha}$ is a gauge-invariant,
first-order quantity according to the Stewart-Walker Lemma~\citep{Stewart_Walker_1974}.
Moreover, the tilting angle $\beta$ associated with the isotropic
frame transformation between the comoving and the barred frames also
vanishes in the background and is a gauge-invariant, first-order quantity
with respect to the background (cf. Appendix~\ref{Appendix:sec_isotropic_frame_transformations}).
Then, at first-order, that is, disregarding terms
with products of multiple first-order quantities with respect to the
equilibrium background spacetime, Eq.~(\ref{eq:Particle_density_conservation_bar_alt})
reads
\begin{equation}
\frac{\nabla_{\alpha}\bar{n}^{\alpha}}{\bar{n}}=-\beta\left(\mathcal{A}_{0}+\phi_{0}+\frac{\widehat{n}_{0}}{n_{0}}\right)-\bar{e}^{\,\alpha}\nabla_{\alpha}\beta\,,
\end{equation}
where $\widehat{n}_{0}=e_{0}^{\alpha}\nabla_{\alpha}n_{0}$. Using
Eq.~(\ref{eq:Frame_transformations_beta_heatFlow_linear_order}),
we find the following relation valid at linear level:
\begin{equation}
\frac{\nabla_{\alpha}\bar{n}^{\alpha}}{\bar{n}}=\left(\phi_{0}+2\mathcal{A}_{0}+\frac{\widehat{n}_{0}}{n_{0}}-\frac{\hat{\mu}_{0}}{\mu_{0}+p_{0}}\right)\frac{\bar{Q}}{\mu_{0}+p_{0}}+\frac{\widehat{\bar{Q}}}{\mu_{0}+p_{0}}\,,\label{eq:Heat_transfer_fractional_variation_particle density_vector_intermediate}
\end{equation}
where $\widehat{\mu}_{0}=e_{0}^{\alpha}\nabla_{\alpha}\mu_{0}$ and
$\widehat{\bar{Q}}=\bar{e}^{\,\alpha}\nabla_{\alpha}\bar{Q}$.

Last, we analyze the quantity $\widehat{n}_{0}/n_{0}$ in terms of
the energy density and pressure of the equilibrium fluid. Using the
definition of specific volume in the case of the equilibrium fluid:
$v_{0}=1/\left(\mu_{N}\right)_{0}$, for the
background spacetime the total relativistic energy density is given by $\mu_{0}=\left(\mu_{N}\right)_{0}\left(1+\varepsilon_{0}\right)$,
where $d\varepsilon_{0}=-p_{0}dv_{0}$. Then,

\begin{equation}
\frac{\widehat{n}_{0}}{n_{0}}=\frac{\widehat{\mu}_{0}}{\mu_{0}+p_{0}}\,,
\end{equation}
and Eq.~(\ref{eq:Heat_transfer_fractional_variation_particle density_vector_intermediate})
simplifies to,
\begin{equation}
\frac{1}{\bar{n}}\nabla_{\alpha}\bar{n}^{\alpha}=\frac{\phi_{0}+2\mathcal{A}_{0}}{\mu_{0}+p_{0}}\bar{Q}+\frac{\widehat{\bar{Q}}}{\mu_{0}+p_{0}}\,.
\label{eq:Heat_transfer_fractional_variation_particle density_vector_final}
\end{equation}
Substituting Eq.~(\ref{eq:Heat_transfer_fractional_variation_particle density_vector_final})
in Eq.~(\ref{eq:Heat_transfer_dot_energy_density_static_frame}),
at first order,  we find
\begin{equation}
\dot{\bar{\mu}}=-\left(\mu_{0}+p_{0}\right)\bar{\theta}-\left(\phi_{0}+2\mathcal{A}_{0}\right)\bar{Q}-\widehat{\bar{Q}}+\bar{\mu}_{N}\overline{T\vphantom{\bar{t}}}\dot{\bar{S}}\,.
\label{eq:Conservation_energy_linear_order}
\end{equation}

\subsection{Second order nature of the heat flow}

Now, the general energy conservation law $\bar{u}_{\beta}\nabla_{\alpha}\bar{T}^{\alpha\beta}=0$ for a spherically symmetric spacetime, in the language of the 1+1+2 formalism reads~\cite{Luz_Carloni_2024a}
\begin{equation}
\dot{\bar{\mu}}=-\left(\bar{\mu}+\bar{p}\right)\bar{\theta}-\left(\bar{\phi}+2\bar{\mathcal{A}}\right)\bar{Q}-\widehat{\bar{Q}}-\frac{3}{2}\bar{\Pi}\bar{\Sigma}\,.\label{eq:Conservation_energy_1p1p2_general}
\end{equation}
Taking into account the discussion in Appendix~\ref{Appendix:sec_isotropic_frame_transformations}, at a linear perturbative level with respect to the background spacetime, Eq.~\eqref{eq:Conservation_energy_1p1p2_general} reads
\begin{equation}
\dot{\bar{\mu}}=-\left(\mu_0+p_0\right)\bar{\theta}-\left(\phi_0+2\mathcal{A}_0\right)\bar{Q}-\widehat{\bar{Q}}+\text{higher-order terms}\,.\label{eq:Conservation_energy_1p1p2_general_linear_order}
\end{equation}

Comparing Eqs.~\eqref{eq:Conservation_energy_linear_order} and 
\eqref{eq:Conservation_energy_1p1p2_general_linear_order}, we see that the term 
$\overline{T\vphantom{\bar{t}}}\dot{\bar{S}}$ is of higher perturbative order. Therefore, at linear level, there is no entropy change of the volume elements of the fluid along the world lines of the static observer: $\dot{\bar{S}}=0$. This result and the assumption of a quasi-static
thermodynamic evolution implies that the processes within the volume elements of the fluid are reversible. Therefore, at linear level,
heat transfer verifies
\begin{equation}
\int_{c}\delta{\bar{\mathsf{Q}}}=\int_{\tau_i}^{\tau_f}\overline{T\vphantom{\bar{t}}}\dot{\bar{S}}\,d\tau=0\,,
\label{eq:Heat_transfer_no_heat_exchange}
\end{equation}
leading us to conclude that along the world-line $c$ of a static observer there is no
exchange of energy as heat of an infinitesimal volume element. Therefore, a reversible adiabatic process in the comoving frame
is also reversible and adiabatic in the static frame if the two frames
are related by an isotropic frame transformation that is first-order
with respect to a background spacetime.
This conclusion is in line with the discussion in Refs.~\cite{Chandrasekhar_1964_PRL_b, Pauli_book}, confirming that if the thermodynamic processes are carried out quasi-statically, the correction to the thermodynamic behavior due to the motion between two frames is a higher than linear-order effect.

In addition, Eq.~(\ref{eq:Heat_transfer_fractional_variation_particle density_vector_final})
explicitly characterizes the origin of the momentum flux term in the
stress-energy tensor in the static frame, Eq.~\eqref{Isotropic_frame_transformations_eq:stress_energy_tensor_transformed_perfect_fluid_simplified}: $\bar{Q}$ and its derivative
account for the fractional divergence of the particle density flux
vector along the $\bar{e}$ direction, that is, the momentum flux
terms account for the net particle flux entering and leaving the volume
element associated with the static observer. However, Eq.~(\ref{eq:Heat_transfer_no_heat_exchange})
asserts that at first order, there is no change along $\bar{u}$ of
the energy of the system as heat. Hence, these terms do not represent
dissipative heat fluxes within the fluid. To clarify, the $Q$ terms
in the stress-energy tensor of a fluid represent only dissipative
effects in the fluid's rest frame. In other frames, these terms also
characterize energy carried by matter fluxes between volume elements
of fluid, even if energy is exchanged as work.

\subsection{Entropy and temperature of the perturbed fluid}

In the previous subsection, we have discussed that for an adiabatic, quasi-static process
in the comoving frame, a radially static observer will still measure
momentum fluxes within the fluid. However, we have proved that in
first-order perturbation theory with respect to a background spacetime,
those fluxes do not represent dissipative effects, and the process
is reversible. This implies that, in the considered perturbative regime,
the momentum fluxes do not lead to a change in the total entropy of
the system. Nonetheless, it is natural to ask how these fluxes are
related with a perceived entropy flux between infinitesimal elements
of volume of the fluid and if it is possible to find the local value for the
temperature of the fluid in the static, non-inertial frame.
Considering the local-thermodynamic equilibrium ansatz,
we can understand these effects by evaluating how an isotropic frame
transformation of the type described in 
Appendix~\ref{Appendix:sec_isotropic_frame_transformations},
which are first-order with respect to a static background, affects
the description of the ``entropy flow'' in and out of the infinitesimal
volume elements of the perturbed fluid defined by the radially static
observers.
Then, let $S^{\alpha}$ represent, formally, the entropy flow density 4-vector in the
frame comoving with the volume elements of the matter fluid, with
4-velocity $u$, such that
\begin{equation}
S^{\alpha}=m_{N}SN^{\alpha}=m_{N}Snu^{\alpha}\,,
\end{equation}
where $S$ denotes the specific entropy and $N^{\alpha}$ represents
the particle 4-current density vector with respect to a comoving observer,
as defined in Eq.~(\ref{eq:Particle_density_vector_general_definition}).
Consider an isotropic frame transformation~(\ref{eq:Frame_transformations_ue_noBar_to_Bar}),
and let the barred frame be associated with a radially static observer
with 4-velocity $\bar{u}$. Using Eqs.~(\ref{eq:Particle_density_vector_Bar_general_definition})
and (\ref{eq:Particle_density_Bar_NoBar_relations}), the components
of the entropy flow density 4-vector transform as
\begin{equation}
S^{\alpha}=m_{N}Snu^{\alpha}\to\bar{S}^{\,\alpha}=m_{N}S\left(n\bar{u}^{\alpha}\cosh\beta\right)-m_{N}S\left(n\bar{e}^{\,\alpha}\sinh\beta\right)\,.\label{eq:Heat_transfer_entropy_flow_transformation}
\end{equation}
In particular, as expected, $\bar{S}^{\,\alpha}=m_{N}S\bar{N}^{\alpha}$.
The first term in the right-hand side of Eq.~(\ref{eq:Heat_transfer_entropy_flow_transformation})
is called the ``convection term,'' and it accounts for the entropy
density flow carried along the direction of $\bar{u}$. The second
term, called the ``conduction and diffusion term'', accounts for
the entropy carried by the matter flux entering and leaving the volume
element along the $\bar{e}$ direction. Indeed, to confirm this interpretation,
we will explicitly relate the momentum transfer term $\bar{Q}$ with
the diffusion term in the considered perturbative setup.

For simplicity, let
\begin{equation}
\bar{s}^{\,\alpha}:=-m_{N}S\left(n\bar{e}^{\,\alpha}\sinh\beta\right)=-m_{N}S\bar{n}^{\alpha}\,,\label{eq:Heat_transfer_entropy_diffusion_static_frame}
\end{equation}
represent the conduction and diffusion term in the static frame. As explained previously,
in the equilibrium background spacetime, observers comoving with the
fluid are also radially static, therefore $\bar{s}^{\alpha}$ is zero
in the background and is a gauge-invariant quantity according with
the Stewart-Walker Lemma. Then, using Eq.~(\ref{eq:Frame_transformations_beta_heatFlow_linear_order}),
valid in first-order perturbation theory, Eq.~(\ref{eq:Heat_transfer_entropy_diffusion_static_frame})
reduces to
\begin{equation}
\bar{s}^{\,\alpha}=\frac{m_{N}n_{0}S_{0}}{\mu_{0}+p_{0}}\bar{Q}\bar{e}^{\,\alpha}\,.\label{eq:Heat_transfer_entropy_diffusion_static_linear_order}
\end{equation}
On the other hand, a general diffusion term that depends linearly
on the heat flux density can be written as
\begin{equation}
\bar{s}^{\,\alpha}=\frac{\eta\bar{Q}\bar{e}^{\,\alpha}}{\vphantom{\bar{\bar{Q}}}\overline{T\vphantom{\bar{t}}}}\,,\label{Heat_transfer_entropy_diffusion_static_linear_order_coefficient}
\end{equation}
where $\eta$ is a thermodynamic coefficient
and, as in the previous section, $\overline{T\vphantom{\bar{t}}}$ is the local temperature.
Comparing Eqs.~(\ref{eq:Heat_transfer_entropy_diffusion_static_linear_order})
and (\ref{Heat_transfer_entropy_diffusion_static_linear_order_coefficient})
implies
\begin{equation}
\frac{\eta}{\vphantom{\bar{\bar{Q}}}\overline{T\vphantom{\bar{t}}}}=\frac{m_{N}n_{0}S_{0}}{\mu_{0}+p_{0}}\,.\label{eq:Heat_transfer_entropy_Temperature}
\end{equation}
To interpret this result, notice that, in the geometrized units system, we can set $\eta=1$ and Eq.~\eqref{eq:Heat_transfer_entropy_Temperature} takes the familiar form for the inverse temperature found in equilibrium thermodynamics of relativistic continuous media for fluids composed of a single species (cf., e.g., Ref.~\cite{Israel_Stewart_1979}). Therefore, by imposing the local-thermodynamic equilibrium ansatz,
Eq.~\eqref{eq:Heat_transfer_entropy_Temperature} asserts that,
up to linear order, the entropy flux density, $\bar{s}^{\,\alpha}$, and the momentum flux density, $\bar{Q}\bar{e}^{\,\alpha}$, are related simply by the value of the local temperature of the equilibrium fluid.

\section{Adiabatic radial perturbations\label{sec:AdiabaticRadialPerturbations}}

In the previous section, we have established that to characterize a
perfect fluid in frames other than in a frame locally comoving with
the fluid, the stress-energy density has to contain non-diagonal terms
to account for the momentum transfer in and out of the infinitesimal
volume elements of the matter fluid. This implies that the description
of the fluid becomes more complex in those other frames. However,
as we will see, this freedom allows us to pick a non-inertial frame
in which the dynamical description of the fluid and the geometry of
the spacetime can be greatly simplified, such that under certain general
conditions, the problem of the determination of the behavior of the perturbations can be treated using standard analytic methods
efficiently.

In Ref.~\cite{Luz_Carloni_2024a}, a general perturbation theory was developed that is manifestly covariant and identification gauge invariant. This theory can then be applied to study perturbations of static, compact, spherically symmetric solutions of the theory of general relativity. For simplicity, we call these types of solutions ``stars'', although these are suitable to model any
static, self-gravitating relativistic matter distributions. 
The new perturbation scheme was constructed using the so-called 1+1+2 covariant formalism, introduced in Ref~\citep{Clarkson_2007}. However, for clarity, in this and subsequent sections, we will consider a particular coordinate system and write the covariant variables in terms of the metric and its derivatives.

\subsection{The equilibrium spacetime and the perturbation variables}\label{Eq&PertVar}

Consider the Einstein field equations (EFE)

\begin{equation}
R_{\alpha\beta}-\frac{1}{2}g_{\alpha\beta}R=T_{\alpha\beta}\,,
\end{equation}
where $R_{\alpha\beta}$ are the components of the
Ricci tensor, $R$ the Ricci scalar, and $T_{\alpha\beta}$ represents the components of
the metric stress-energy tensor. For simplicity, we have set the cosmological
constant to zero.

Let the equilibrium background spacetime be a spatially compact, static,
spherically symmetric solution of the EFE with a perfect fluid source,
such that
\begin{equation}
T_{\alpha\beta}=\left(\mu_{0}+p_{0}\right)\left(u_{0}\right)_{\alpha}\left(u_{0}\right)_{\alpha}+p_{0}\left(g_{0}\right)_{\alpha\beta}\,,
\end{equation}
where $u_{0}$ is the 4-velocity of an observer locally comoving with the volume elements of the
fluid, $\left(g_{0}\right)_{\alpha\beta}$ are the components of the
metric tensor in some local coordinate system, $\mu_{0}$ represents the energy density and $p_{0}$ the isotropic pressure of
the matter fluid. As we have done in the previous section and from
hereon, we will use the ``0'' to explicitly refer to quantities
of the equilibrium spacetime. Generically, in the  coordinates $\left( t, r, \psi, \varphi \right)$ defined by a static observer at spatial infinity, this type of solutions
can be characterized by a line element of the form
\begin{equation}
ds_{0}^{2}=-\left(g_0\right)_{tt}\,dt^{2}+\left(g_0\right)_{rr}\,dr^{2}+
r^{2} \left( d \psi^2+\sin^2\psi\,d\varphi^2 \right)\,,
\label{eq:general_static_line_element}
\end{equation}
where the metric coefficients
$\left(g_0\right)_{tt}$ and $\left(g_0\right)_{rr}$ are assumed to be functions solely of $r$.

We will consider the equilibrium spacetime manifold to be composed of two solutions 
of the EFE of the type \eqref{eq:general_static_line_element}. The first one, representing the exterior of the star, is a regular branch of the
Schwarzschild solution, the second one, modeling the interior of the star, is a static spatially
compact solution with a perfect fluid matter source. The two solutions are smoothly matched at $r=r_{b}$  (the boundary of the star) to each other at a common timelike hypersurface via the standard Israel junction formalism.

For the line element~(\ref{eq:general_static_line_element}), we
can write the 1+1+2 covariant scalar functions directly in terms of
the metric components and its derivatives as
\begin{equation}
\begin{aligned}\phi_{0} & =\frac{2}{r\sqrt{\left(g_0\right)_{rr}}}\,,\\
\mathcal{A}_{0} & =\frac{1}{2\left(g_0\right)_{tt}\sqrt{\left(g_0\right)_{rr}}}\frac{d\left(g_0\right)_{tt}}{dr}\,,
\end{aligned}
\label{eq:Background_phi_A_E}
\end{equation}
where $\phi_{0}$ characterizes the spatial expansion of the normalized
radial gradient vector field, and $\mathcal{A}_{0}$ is the radial
component of the 4-acceleration of an observer locally comoving with the volume elements of the matter fluid.

Now, in Ref.~\citep{Stewart_Walker_1974}, necessary and sufficient conditions were established for a quantity to be independent 
of the choice of diffeomorphism between an equilibrium and a perturbed
spacetime, i.e., to be identification gauge-invariant. Therefore, by identifying an appropriate closed set of gauge-invariant perturbation variables, we can
characterize 
the perturbed spacetime unambiguously.
In that regard, consider the energy density, $\bar{\mu}$, and pressure, $\bar{p}$,
of the perturbed fluid measured in the radially static frame. These
are not gauge-invariant quantities, however, given that the background
spacetime is assumed static, their proper time derivatives vanish
in the background, hence, are gauge-invariant and can be used to characterize the perturbed spacetime.
Indeed, to describe the perturbations, we will consider the variables
\begin{equation}
\begin{aligned}\mathsf{m} & :=\dot{\bar{\mu}}\,, & \mathsf{p} & :=\dot{\bar{p}}\,, & \mathsf{A} & :=\dot{\bar{\mathcal{A}}}\,, & \mathsf{F} & :=\dot{\bar{\phi}}\,, & \mathsf{E} & :=\dot{\bar{\mathcal{E}}}\,,\end{aligned}
\label{eq:GI_dot_derivatives_definition}
\end{equation}
where the general definitions of $\mathcal{A}$, $\phi$ and $\mathcal{E}$
can be found in Appendix~\ref{Appendix:sec_1p1p2_decomposition}, and,
as in the previous section, ``dot'' represents derivatives along
the world-lines of the radially static observers, that is, derivatives
taken with respect to the proper time of those observers. In addition
to the quantities in Eq.~(\ref{eq:GI_dot_derivatives_definition}),
we will also consider the expansion scalar, $\bar{\theta}$, and the
nontrivial radial component of the shear tensor, $\bar{\Sigma}$,
both associated with the local congruence formed by the world-lines
of the radially static observers since those quantities vanish identically in
the background spacetime. For the radially static frame, as we have
discussed in the previous section, we also need to include a momentum
transfer term $\bar{Q}$ in the fluid's stress-energy tensor. Then, in a radially static frame,
adiabatic, radial perturbations of perfect fluid stars can be, independently of the choice of gauge, fully characterized by the variables $\left\{ \mathsf{m},\mathsf{p},\bar{Q},\mathsf{A},\mathsf{F},\mathsf{E},\bar{\theta},\bar{\Sigma}\right\} $.

\subsection{Harmonic decomposition}

The perturbation equations are found by linearizing the EFE. The linearization procedure dramatically simplifies the field
equations, however, these can be further simplified by considering
the exact symmetries of the equilibrium spacetime to transform the
linearized system of partial differential equations in a system
of ordinary differential equations.

In this article, we will consider the background spacetime to be spatially
compact, static, and with spherical symmetry. Then, at linear perturbation order, it is possible to express a covariantly defined scalar perturbation
variable, $\chi$, in terms of  the eigenfunctions $e^{i\upsilon\tau}$ of the Laplace operator in $\mathbb{R}$,
of the background spacetime, and the Spherical Harmonics,
$Y_{\ell m}$:
\begin{equation}
\chi=\sum_{\upsilon}\left(\sum_{\ell=0}^{+\infty}\sum_{m=-\ell}^{\ell}\Psi_{\chi}^{\left(\upsilon,\ell\right)}Y_{\ell m}\right)e^{i\upsilon\tau}\,,
\label{eq:general_harmonic_expansion}
\end{equation}
where $\tau$ is the proper time of the comoving observer in the background and $\upsilon$ parameterizes the associated eigenfrequencies, $\sum_{\upsilon}$ can stand for a discrete sum or an integral
in $\upsilon$, depending on the boundary conditions of the problem,
and the coefficients $\Psi_{\chi}^{\left(\upsilon,\ell\right)}$ are functions of $r$ only.

In this article, we will consider isotropic perturbations, therefore dipole and higher-order
angular multipoles are zero as those would select preferred directions in the system. Then, in the expansion~\eqref{eq:general_harmonic_expansion} we can disregard all coefficients $\Psi_{\chi}^{\left(\upsilon,\ell\right)}$
with $\ell\geq1$. Moreover, provided sufficient regularity for the  background
spacetime and the perturbed fluid, the squared eigenfrequencies, $\upsilon^{2}$,
are countable, real, simple, have a minimum, and are unbounded from
above~\cite{Luz_Carloni_2024a}. Therefore, in the considered setup, any first-order, gauge-invariant scalar quantity, $\chi$, can be decomposed
as
\begin{equation}
\chi=\sum_{\upsilon^{2}=\left\{ \upsilon_{0}^{2},\upsilon_{1}^{2},...\right\} }\Psi_{\chi}^{\left(\upsilon\right)}(r)Y_{00}\,e^{i\upsilon\tau}\,,
\label{eq:Radial_Adiabatic_general_harmonic_decompostion_upsilon}
\end{equation}
where we have dropped the superscript $\ell$.

The expansion above was done employing the proper time, $\tau$, of an observer locally
comoving with the equilibrium fluid. However, it is more advantageous in some cases to express it by 
means of the time coordinate $t$. The eigenfrequencies associated with each of these time coordinates are connected by 
the relation~\cite{Betschart_Clarkson_2004}
\begin{equation}
\upsilon\left(r\right)=\lambda\exp\left(\int_{+\infty}^{r}-\frac{2\mathcal{A}_{0}}{x\phi_{0}}dx\right)=\frac{\lambda}{\sqrt{\left(g_{0}\right)_{tt}}}\,,
\label{eq:Radial_Adiabatic_eigenfrequencies_v-lambda_relation}
\end{equation}
where the constant $\lambda$ represents the value of an eigenfrequency measured by the observer at spatial infinity.
Thus, in terms of $t$, a gauge-invariant, first-order scalar quantity $\chi$ can be equivalently given by
\begin{equation}
\chi=\sum_{\lambda^{2}=\left\{ \lambda_{0}^{2},\lambda_{1}^{2},...\right\} }\Psi_{\chi}^{\left(\lambda\right)}Y_{00}\,e^{i\lambda t}\,.
\end{equation}

\subsection{Gauge-invariant equation of state and perturbation equations}

The dynamical evolution of the adiabatically,
radially perturbed spacetime is characterized by a system of equations for the perturbation variables \eqref{eq:GI_dot_derivatives_definition}. 
Nonetheless, this system is not closed and requires additional information related to the thermodynamics of the fluid source in the form of an equation
of state. In what follows, we will assume that in the
local rest frame of the perturbed fluid, the pressure, and the energy density are related by
\begin{equation}
p=f\left(\mu\right)\,,
\end{equation}
i.e., a barotropic equation of state. Here $f$ is a generic twice differentiable function 
defined in an open neighborhood of $\mu_{0}$.
With these assumptions, $f'\left(\mu_{0}\right)$ can be associated to the square of
the adiabatic speed of sound of  fluid as measured by a comoving
observer. In line with the prescriptions of a physically meaningful background, we will assume that $f'\left(\mu_{0}\right)$ is non-vanishing
in the interior of the perturbed star. The equation of state above translates in the comoving frame in a similar relation for the perturbation variables $\mathsf{m}$ and $\mathsf{p}$. However, as mentioned in Section \ref{sec:Thermodynamic-description}, in the static frame the fluid is no longer perceived to be perfect, therefore the equation of state is modified, as proven in~\cite{Luz_Carloni_2024a}, into the relation  
\begin{equation}
\mathsf{m}=\frac{1}{f'\left(\mu_{0}\right)}\mathsf{p}-\frac{r\phi_{0}}{2\left(\mu_{0}+p_{0}\right)}\left(\frac{d\mu_{0}}{dr}-\frac{1}{f'\left(\mu_{0}\right)}\frac{dp_{0}}{dr}\right)Q\,,
\label{eq:Static_Radial_Adiabatic_corrected_eos}
\end{equation}
where we have dropped the overline for simplicity's sake. In the remaining of the article
all first-order quantities are to be considered those measured in
the frame of radially static observers within the star.
Moreover, notice that if the equilibrium fluid is characterized by a barotropic equation of state equal to that of the perturbed fluid, the second term in the RHS of Eq.~\eqref{eq:Static_Radial_Adiabatic_corrected_eos} vanishes. We also remark that we have not imposed any direct constraints on the particular type of the equation of state of the equilibrium fluid.

Gathering the previous results, we are now in a position to write the
perturbation equations. In the coordinate system, $\left(t,r, \psi, \varphi\right)$,
and considering the harmonic decomposition~\eqref{eq:Radial_Adiabatic_general_harmonic_decompostion_upsilon}, the nontrivial harmonic coefficients
verify the following system of equations~\cite{Luz_Carloni_2024a}
\begin{equation}
\begin{aligned}
	\frac{d\Psi_{\mathsf{p}}^{\left(\upsilon\right)}}{dr} = &	\frac{2}{r\phi_0}
	\left[\frac{\mu_{0}+p_{0}}{\phi_{0}}\left(\frac{1}{2}\phi_{0}+2\mathcal{A}_{0}\right)+\frac{\mathcal{A}_{0}^{2}}{f'\left(\mu_{0}\right)}+\frac{r\mathcal{A}_{0}\phi_{0}}{2\left(\mu_{0}+p_{0}\right)}\frac{d\mu_{0}}{dr}+\upsilon^{2}\right]\Psi_{Q}^{\left(\upsilon\right)}\\
 & -\frac{2}{r\phi_0}\left[\frac{\mu_{0}+p_{0}}{\phi_{0}}+\left(2+\frac{1}{f'\left(\mu_{0}\right)}\right)\mathcal{A}_{0}\right]\Psi_{\mathsf{p}}^{\left(\upsilon\right)}\,,
\end{aligned}
\label{eq:Static_Radial_Adiabatic_pdot_hat}
\end{equation}
\begin{equation}
\begin{aligned}
	\frac{d\Psi_{Q}^{\left(\upsilon\right)}}{dr}= & \frac{2}{r\phi_0}\left[\frac{\mathcal{A}_{0}}{f'\left(\mu_{0}\right)}+\frac{r\phi_{0}}{2\left(\mu_{0}+p_{0}\right)}\frac{d\mu_{0}}{dr}+\frac{\mu_{0}+p_{0}}{\phi_{0}}-\phi_{0}-2\mathcal{A}_{0}\right]\Psi_{Q}^{\left(\upsilon\right)} -\frac{2}{r\phi_0 f'\left(\mu_{0}\right)}\Psi_{\mathsf{p}}^{\left(\upsilon\right)}\,,
\end{aligned}
\label{eq:Static_Radial_Adiabatic_Q_hat}
\end{equation}
and
\begin{equation}
\begin{aligned}\Psi_{\mathsf{m}}^{\left(\upsilon\right)} & =\frac{1}{f'\left(\mu_{0}\right)}\Psi_{\mathsf{p}}^{\left(\upsilon\right)}-\left(\frac{\mathcal{A}_{0}}{f'\left(\mu_{0}\right)}+\frac{r\phi_{0}}{2\left(\mu_{0}+p_{0}\right)}\frac{d\mu_{0}}{dr}\right)\Psi_{Q}^{\left(\upsilon\right)}\,,\\
\Psi_{\mathsf{A}}^{\left(\upsilon\right)} & =\frac{1}{\phi_{0}}\left[\Psi_{\mathsf{p}}^{\left(\upsilon\right)}-\left(\frac{1}{2}\phi_{0}+\mathcal{A}_{0}\right)\Psi_{Q}^{\left(\upsilon\right)}\right]\,,\\
\Psi_{\mathsf{E}}^{\left(\upsilon\right)} & =\frac{1}{2}\phi_{0}\Psi_{Q}^{\left(\upsilon\right)}+\frac{1}{3f'\left(\mu_{0}\right)}\Psi_{\mathsf{p}}^{\left(\upsilon\right)}\,,\\
\Psi_{\theta}^{\left(\upsilon\right)} & =-\frac{1}{\phi_{0}}\Psi_{Q}^{\left(\upsilon\right)}\,,\\
\Psi_{\Sigma}^{\left(\upsilon\right)} & =\frac{2}{3}\Psi_{\theta}^{\left(\upsilon\right)}\,,\\
\Psi_{\mathsf{F}}^{\left(\upsilon\right)} & =\Psi_{Q}^{\left(\upsilon\right)}\,,
\end{aligned}
\label{eq:Static_Radial_Adiabatic_constraint_eqs}
\end{equation}
where the second to last equation is the kinematical requirement for the constancy of the circumferential radius coordinate of the static observer.

To solve the system of differential equations above, we must impose boundary conditions. Given the particular physical setup that we are interested in, we consider that:
\begin{enumerate}[label=(\roman*)]
\item \label{enu:general_boundary_condition_1} at the center of the star, $r=0$, and at the initial instant, the energy density and the
pressure perturbations must be finite;
\item \label{enu:general_boundary_condition_2} there exists a timelike hypersurface at which the interior spacetime and an exterior vacuum Schwarzschild spacetime can be smoothly matched.
\end{enumerate}
As was proven in Ref.~\cite{Luz_Carloni_2024a}, for the static observer, condition~\ref{enu:general_boundary_condition_2}
sets that at the surface of the star, at all times, we must have
\begin{equation}
\left.\mathsf{p}-\mathcal{A}_{0}Q\right|_{\text{boundary}}=0\,,
\end{equation}
therefore,
\begin{equation}
\left.\Psi_{\mathsf{p}}^{\left(\upsilon\right)}-\mathcal{A}_{0}\Psi_{Q}^{\left(\upsilon\right)}\right|_{\text{boundary}}=0\,.
\label{eq:Static_Radial_Adiabatic_boundary_condition_coefficients}
\end{equation}

Before we discuss the solutions of the system~(\ref{eq:Static_Radial_Adiabatic_pdot_hat})--(\ref{eq:Static_Radial_Adiabatic_constraint_eqs}),
we comment on some glaring differences between this system and the one found
for the comoving frame, discussed in Ref.~\cite{Luz_Carloni_2024b}.
In the comoving frame, three master perturbation variables are necessary
to characterize adiabatic, radial perturbations, whereas, in the radially
static frame, the description of adiabatic, radial perturbations is
completely characterized by two master variables: $\mathsf{p}$ and
$Q$. Moreover, to characterize this type of perturbations in the
radially static frame, only information on the square
of the speed of sound of the fluid, $f'$, is required.
This is not surprising. As noted in Ref.~\citep{Vath_Chanmugam_1992},
where the original Chandrasekhar second-order radial pulsation equation
was recast as a first-order coupled system of differential equations,
the new system also only depends on the value of the adiabatic index
of the fluid and not of its derivative. Therefore, the matter model
is completely characterized by the values of the adiabatic index or,
equivalently, by the square of the speed of sound measured in the
fluid's local rest frame.

\subsection{Analytic solutions}

Given the background spacetime and the equation of state of the
perturbed fluid, Eqs.~(\ref{eq:Static_Radial_Adiabatic_pdot_hat}) and (\ref{eq:Static_Radial_Adiabatic_Q_hat}),
together with conditions~\ref{enu:general_boundary_condition_1}
and \ref{enu:general_boundary_condition_2}, completely describe adiabatic, radial perturbations of a star composed of a perfect fluid,
from the point of view of a radially static observer within the star. This type of systems of ordinary differential equations is well-suited to employ numerical methods to find approximate solutions. Notwithstanding, we have shown in~\cite{Luz_Carloni_2024a} that it is also possible to find analytic solutions in the form of power series,
using methods borrowed from the standard theory of systems of linear ordinary
differential equations. In that regard,  let the following extra conditions:
\begin{enumerate}[label=(\alph*)]
\item \label{enu:regularity_conditions} the equilibrium fluid variables $\mu_{0}$ and
$p_{0}$ verify the
weak energy condition;
\item \label{enu:regularity_conditions2} the equilibrium star is a solution of the TOV
equations for real analytic, nontrivial energy density, and isotropic pressure functions; 
\item  \label{enu:regularity_conditions3} the function $f'$, representing the square of the adiabatic speed of sound of the perturbed fluid in the comoving
frame, is positive and real analytic within the perturbed star.
\end{enumerate}
Condition~\ref{enu:regularity_conditions2}, in particular, is a very strong restriction to the type of spacetimes that can be considered, especially taking into account the nontrivial equations of state that are characteristic of nuclear matter. Serendipitously, to our knowledge, all known classical exact solutions for compact astrophysical objects have this property at least in a neighborhood of 
the center of the star.
Indeed, if the equilibrium spacetime is a solution of the EFE with real analytic $\mu_0$ and $p_0$ matter variables, such that both their power series centered at $r=0$ have a nonzero radius of convergence, the technique shown below can be used to characterize the perturbed spacetime, under adiabatic, radial perturbations. Nonetheless, we remark that if the radius of convergence of any of the power series is smaller than the radius of the equilibrium star, the validity of the solutions presented below is only guaranteed for points whose circumferential radius coordinate is smaller than that value of the radius of convergence.

Continuing, considering the regularity conditions above, Eq.~(\ref{eq:Background_phi_A_E}) can be rewritten as
\begin{equation}
\begin{aligned}\phi_{0} & =\frac{2}{r}\sqrt{1-\frac{2M\left(r\right)}{r}}\,,\\
\mathcal{A}_{0}\phi_{0} & =p_{0}+\frac{2M\left(r\right)}{r^{3}}\,,
\end{aligned}
\end{equation}
where
\begin{equation}
M\left(r\right):=\frac{1}{2}\int_{0}^{r}\mu_{0}x^{2}dx\,,
\end{equation}
is dubbed the mass function. Then, conditions~\ref{enu:regularity_conditions} and~\ref{enu:regularity_conditions2} for the matter variables, imply that
the function $\mathcal{A}_{0}$ is real analytic at all points within the star, and the function $\phi_{0}$ is real analytic within the star except at $r=0$ where it has 
a singular point. As a consequence, Eqs.~\eqref{eq:Static_Radial_Adiabatic_pdot_hat} and \eqref{eq:Static_Radial_Adiabatic_Q_hat}
constitute a system of ordinary differential equations with real analytic
coefficients around $r=0$ and  a singular point at $r=0$.

Now, we will start by recasting the system of perturbations equations in the following form:
\begin{equation}
\frac{d}{dr}\mathds{W}=\left(r^{-1}\mathds{R}+\Theta\right)\mathds{W}\,,
\label{eq:Matrix_form_perturbation_system}
\end{equation}
with
\begin{equation}
\begin{aligned}\mathds{W} & =\left[\begin{array}{c}
\Psi_{\mathsf{p}}^{\left(\upsilon\right)}\\
\Psi_{Q}^{\left(\upsilon\right)}
\end{array}\right]\,, &  &  & \mathds{R} & =\left[\begin{array}{lr}
0 & 0\\
0 & -2
\end{array}\right]\,,\end{aligned}
\end{equation}
and
\begin{equation}
\Theta=-\frac{2}{r\phi_{0}}\left[\begin{array}{lr}
\frac{\mu_{0}+p_{0}}{\phi_{0}}+2\mathcal{A}_{0}+\frac{\mathcal{A}_{0}}{f'\left(\mu_{0}\right)} & \qquad-\frac{\mu_{0}+p_{0}}{\phi_{0}}\left(\frac{1}{2}\phi_{0}+2\mathcal{A}_{0}\right)-\frac{\mathcal{A}_{0}^{2}}{f'\left(\mu_{0}\right)}-\frac{r\mathcal{A}_{0}\phi_{0}}{2\left(\mu_{0}+p_{0}\right)}\frac{d\mu_{0}}{dr}-\upsilon^{2}\\
\frac{1}{f'\left(\mu_{0}\right)} & 2\mathcal{A}_{0}-\frac{\mathcal{A}_{0}}{f'\left(\mu_{0}\right)}-\frac{r\phi_{0}}{2\left(\mu_{0}+p_{0}\right)}\frac{d\mu_{0}}{dr}-\frac{\mu_{0}+p_{0}}{\phi_{0}}
\end{array}\right]\,.
\end{equation}

The regularity conditions~\ref{enu:regularity_conditions}-\ref{enu:regularity_conditions3} 
imply that $r\phi_{0}$ is positive in the interior of the star. Then, the matrix $\Theta$ is real analytic at $r=0$, with power series
\begin{equation}
	\Theta\left(r\right)=\sum_{n=0}^{+\infty}\Theta_{n}r^{n}\,.
	\label{eq:Theta_matrix_power_series}
\end{equation}
In particular, we conclude that $r=0$ is a regular singular point of the system~\eqref{eq:Matrix_form_perturbation_system}, and we can employ the
methods presented in Ref.~\citep{Coddington_Levinson_Book} to find the solution $\mathds{W}$ in the form of a convergent power series in the neighborhood of $r=0$. Moreover, the method guarantees that this solution series has a radius of convergence equal to the series in \eqref{eq:Theta_matrix_power_series} except, possibly, at the singular point $r=0$.

Now, for a completely general $\Theta$ matrix, the solutions to the system of differential equations~\eqref{eq:Matrix_form_perturbation_system}
can be quite complicated. However, assuming 
conditions~\ref{enu:regularity_conditions}-\ref{enu:regularity_conditions3}, and making use of the TOV equations, it was found, in general, that some entries of the lowest 
order coefficient-matrices  $\Theta_{0}$ and $\Theta_{1}$ vanish. This greatly 
simplifies the general family of solutions of physical interest. Following Ref.~\citep{Coddington_Levinson_Book}, and given the regularity of the background 
spacetime, the general solutions are given by
\begin{equation}
\left[\begin{array}{c}
\Psi_{\mathsf{p}}^{\left(\upsilon\right)}\\
\Psi_{Q}^{\left(\upsilon\right)}
\end{array}\right]=\left[\begin{array}{cc}
-\frac{1}{r}\left(\Theta_{0}\right)_{12} & \quad1\\
\frac{1}{r^{2}} & 0
\end{array}\right]\mathds{P}_{\mathds{W}}\left[\begin{array}{c}
c_{1}\\
c_{2}
\end{array}\right]\,,
\label{eq:Radial_Adiabatic_static_matrix_system_solutions}
\end{equation}
where $c_{1}$ and $c_{2}$ are integration constants. In the above expression, we have introduced the compact notation $\left(\Theta_{n}\right)_{ij}$ to represent $ij$-element of the $n$th order matrix-coefficient of the power
series of $\Theta$. $\mathds{P}_{\mathds{W}}$ represents a real analytic matrix represented by the series
\begin{equation}
\mathds{P}_{\mathds{W}}\left(r\right)=\sum_{n=0}^{+\infty}\mathds{P}_{n}r^{n}\,,
\end{equation}
where
\begin{equation}
\begin{aligned}\mathds{P}_{0} & =\mathds{I}_{2}\,,\\
\mathds{P}_{k} & =\frac{1}{k}\sum_{j=0}^{k-1}\mathds{A}_{k-1-j}\mathds{P}_{j}\,,\quad\text{for }k\geq1\,,
\end{aligned}
\label{eq:Radial_Adiabatic_static_matrix_system_recurrence_relation}
\end{equation}
$\mathds{I}_{2}$ is the $2\times2$ identity matrix,  and $\mathds{A}_{n}$ is the $n$th order Maclaurin coefficient of the series expansion in $r$
of the matrix
\begin{equation}
\mathds{A}=\left[\begin{array}{lr}
\Theta_{22}-r\left(\Theta_{0}\right)_{12}\Theta_{21} & r^{2}\Theta_{21}\\
\frac{\Theta_{12}-\left(\Theta_{0}\right)_{12}}{r^{2}}+\frac{\left(\Theta_{0}\right)_{12}\left(\Theta_{22}-\Theta_{11}\right)}{r}-\left(\Theta_{0}\right)_{12}^{2}\Theta_{21} & \quad\Theta_{11}+r\left(\Theta_{0}\right)_{12}\Theta_{21}
\end{array}\right]\,.
\end{equation}

To select the physical solutions of  Eqs.~(\ref{eq:Radial_Adiabatic_static_matrix_system_solutions})
and (\ref{eq:Radial_Adiabatic_static_matrix_system_recurrence_relation}),we have to impose the
boundary conditions~\ref{enu:general_boundary_condition_1} and \ref{enu:general_boundary_condition_2}. In that regard,
it is useful to calculate the lower order
coefficients of the power series expansion of solutions matrix $\mathds{W}$. Considering $\mathds{P}_{0}=\mathds{I}_{2}$, the
boundary conditions at $r=0$ imposes that the coefficient
$c_{1}$ must be zero; otherwise the perturbation would diverge at
the center at all times, and we readily find
\begin{equation}
\left[\begin{aligned}\Psi_{\mathsf{p}}^{\left(\upsilon\right)}\\
\Psi_{Q}^{\left(\upsilon\right)}
\end{aligned}
\right]=\left[\begin{aligned}c_{2}+\mathcal{O}\left(r^{2}\right)\\
\mathcal{O}\left(r\right)
\end{aligned}
\right]\,.\label{eq:Radial_Adiabatic_static_coefficients_around_center}
\end{equation}

Once the equilibrium spacetime and the values of the eigenfrequencies
$\upsilon$, or equivalently $\lambda$, are specified, these results allow us to
find real analytic solutions for the perturbation that verify the boundary
conditions.
The previous results, however, do not lead to a closed form
expression to directly compute the values of
$\lambda^2$. Yet, in Ref.~\cite{Luz_Carloni_2024a}, it was
found a constraint to the minimal absolute value of the eigenvalues. Namely, if the boundary conditions~\ref{enu:general_boundary_condition_1}
and \ref{enu:general_boundary_condition_2} are verified, the background spacetime is a $\mathcal{C}^1$ solution of the EFE, the equilibrium fluid verifies the weak energy condition, and $f'\left( \mu_0\right)$ is positive in the interior of the perturbed star,
then, nontrivial $\mathcal{C}^{1}$
solutions of the boundary value problem~\eqref{eq:Static_Radial_Adiabatic_pdot_hat}--\eqref{eq:Static_Radial_Adiabatic_boundary_condition_coefficients}
exist only if
\begin{equation}
\lambda^{2}\max_{r\in\left]0,r_{b}\right[}\left(g_{0}\right)_{tt}>-\max_{r\in\left]0,r_{b}\right[}\left[\frac{\mu_{0}+p_{0}}{\phi_{0}}\left(\frac{1}{2}\phi_{0}+2\mathcal{A}_{0}\right)+\frac{\mathcal{A}_{0}^{2}}{f'\left(\mu_{0}\right)}+\frac{r\phi_{0}\mathcal{A}_{0}}{2\left(\mu_{0}+p_{0}\right)}\frac{d\mu_{0}}{dr}\right]\,.\label{eq:Radial_Adiabatic_comoving_bound_freq}
\end{equation}
This result offers a baseline to determine the values of the eigenfrequencies
of the system numerically.

\section{Perturbations and fundamental eigenfrequencies of classic exact solutions\label{sec:Examples}}

The description of adiabatic, radial perturbations in the static frame
is considerably simpler than that in the comoving frame. This simplification
allows us to study perturbations of a background spacetime more efficiently,
when compared to the algorithm used for the comoving frame. In Ref.~\cite{Luz_Carloni_2024b},
various classical exact solutions of the theory of general relativity
were considered for the equilibrium background spacetime to illustrate
the new perturbation formalism. However, given the complexity of the
perturbation equations, analytic analysis in the comoving frame is
not computationally efficient, to the point where for selected exact
solutions of the theory, the algorithm takes an unreasonable amount
of time to set up the intermediate quantities on an average computer.
In that regard, the
algorithm used to solve the perturbation equations in the static frame
is significantly faster, and we can efficiently determine the eigenfunctions
and the eigenfrequencies of stellar compact objects modeled by 
solutions that have been used in the past to study physically meaningful
scenarios. 

In this section, we will study the properties of adiabatic,
radial perturbations of some well-known self-gravitating equilibria of perfect fluids, namely, 
the Buchdahl I, Heintzmann IIa, Patwardhan-Vaidya
IIa, and Tolman VII solutions. Since the independent
parameter $c_{2}$ simply characterizes the magnitude of a specific
eigenfunction of $\mathsf{p}$ at $r=0$, Eq.~(\ref{eq:Radial_Adiabatic_static_coefficients_around_center}),
without loss of generality in this section, we will consider $c_{2}=1$
for all eigenfunctions. Moreover, we assume that the equation of state
of the perturbed fluid is the same as that of the equilibrium setup.

\begin{table}
\def\arraystretch{1.2}
\centering
\begin{tabular}{|c|c|}
\hline 
Spacetime & Non-trivial metric components\tabularnewline
\hline 
\hline 
Buch1 & $\begin{aligned}\\
\left(g_{0}\right)_{tt}= & A\left[\left(1+Cr^{2}\right)^{\frac{3}{2}}+B\left(5+2Cr^{2}\right)\sqrt{2-Cr^{2}}\right]^{2}\\
\left(g_{0}\right)_{rr}= & \frac{2\left(1+Cr^{2}\right)}{2-Cr^{2}}\\
\\
\end{aligned}
$\tabularnewline
\hline 
Heint IIa & $\begin{aligned}\\
\left(g_{0}\right)_{tt}= & 
A^{2}\left(ar^{2}+1\right)^{3}\\
\left(g_{0}\right)_{rr}= & \left(1-\frac{3ar^{2}\left[c\left(4ar^{2}+1\right)^{-\frac{1}{2}}+1\right]}{2\left(ar^{2}+1\right)}\right)^{-1}\\
\\
\end{aligned}
$\tabularnewline
\hline 
P-V IIa & $\begin{aligned}\\
\left(g_{0}\right)_{tt}= & \left\{ A\cos\left[\frac{1}{2}\text{arcsinh}\left(\frac{b^{2}r^{2}-c}{\sqrt{b^{2}-c^{2}}}\right)+d\right]+B\sin\left[\frac{1}{2}\text{arcsinh}\left(\frac{b^{2}r^{2}-c}{\sqrt{b^{2}-c^{2}}}\right)+d\right]\right\} ^{2}\\
\,\,\left(g_{0}\right)_{rr}= & \left(b^{2}r^{4}-2cr^{2}+1\right)^{-1}\\
\\
\end{aligned}
$\tabularnewline
\hline 
\multirow{1}{*}{\,\,Tolman VII\,\,} & $\begin{aligned}\\
\left(g_{0}\right)_{tt}= & B^{2}\sin^{2}\left[\ln\left(\sqrt{\frac{\sqrt{1-\frac{r^{2}}{R^{2}}+\frac{4r^{4}}{A^{4}}}+\frac{2r^{2}}{A^{2}}-\frac{A^{2}}{4R^{2}}}{C}}\right)\right]\\
\left(g_{0}\right)_{rr}= & \left(1-\frac{r^{2}}{R^{2}}+\frac{4r^{4}}{A^{4}}\right)^{-1}\\
\\
\end{aligned}
$\tabularnewline
\hline 
\end{tabular}

\caption{\label{table:Metric-coefficients}Nontrivial metric coefficients of
solutions of the EFE assuming a line element of the form of Eq.~(\ref{eq:general_static_line_element}). The naming conventions and abbreviations follows those of Ref.~\citep{Delgaty_Lake_1998}.
}
\end{table}

\begin{figure}
\centering\includegraphics[width=1\textwidth]{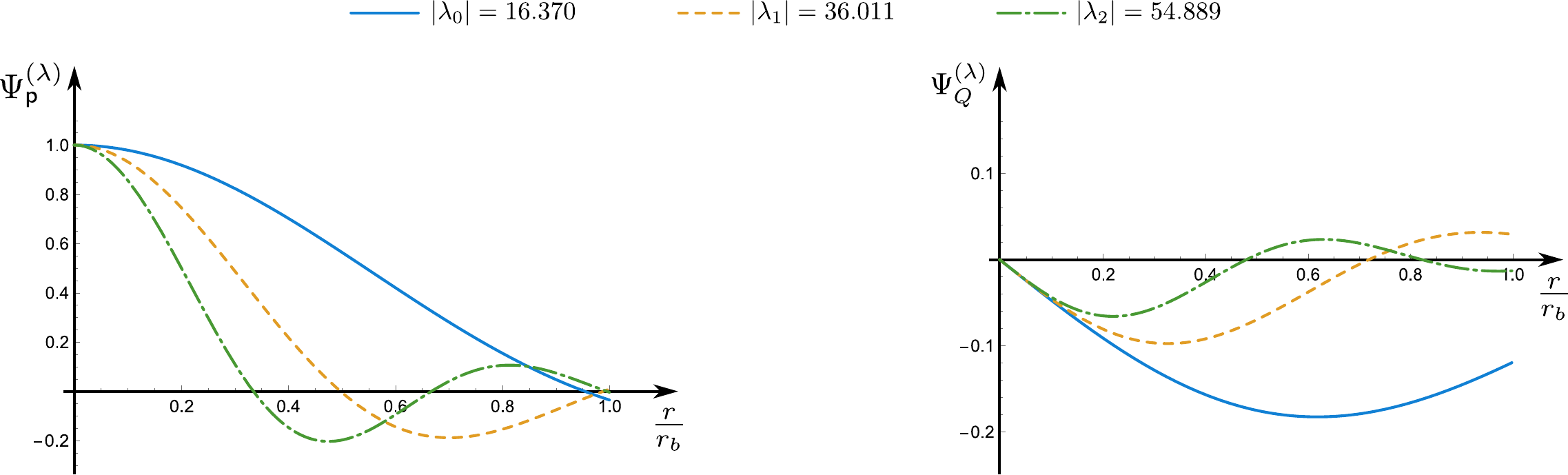}

\caption{\label{fig:Eigenfunctions_Buchdahl_I}
	Behavior of the radial harmonic
	coefficients of the functions $\mathsf{p}$ and $Q$, related with
	the eigenfrequencies in Table~\ref{table:Fundamental-modes}
	for the Buch1 spacetime. It was assumed
	$c_{2}=1$ for all eigenmodes.}
\end{figure}

\begin{figure}
\centering\includegraphics[width=1\textwidth]{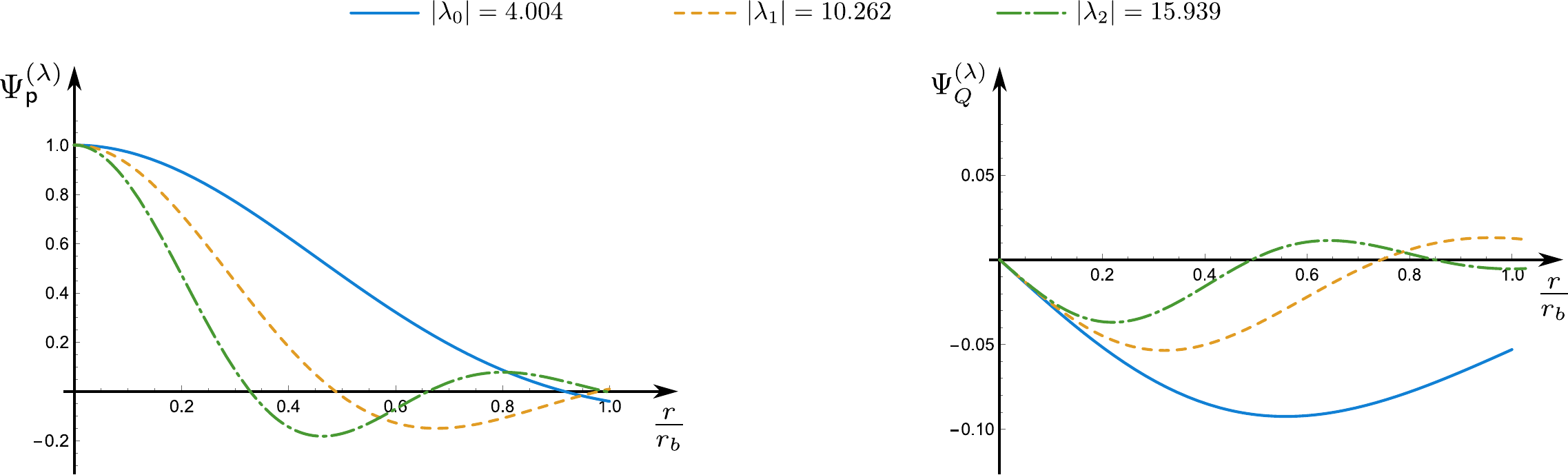}

\caption{\label{fig:Eigenfunctions_Heint_IIa}Behavior of the radial harmonic
	coefficients of the functions $\mathsf{p}$ and $Q$, related with
	the eigenfrequencies in Table~\ref{table:Fundamental-modes}
	for the Heint IIa spacetime. It was assumed
	$c_{2}=1$ for all eigenmodes.}
\end{figure}

\begin{figure}
\centering\includegraphics[width=1\textwidth]{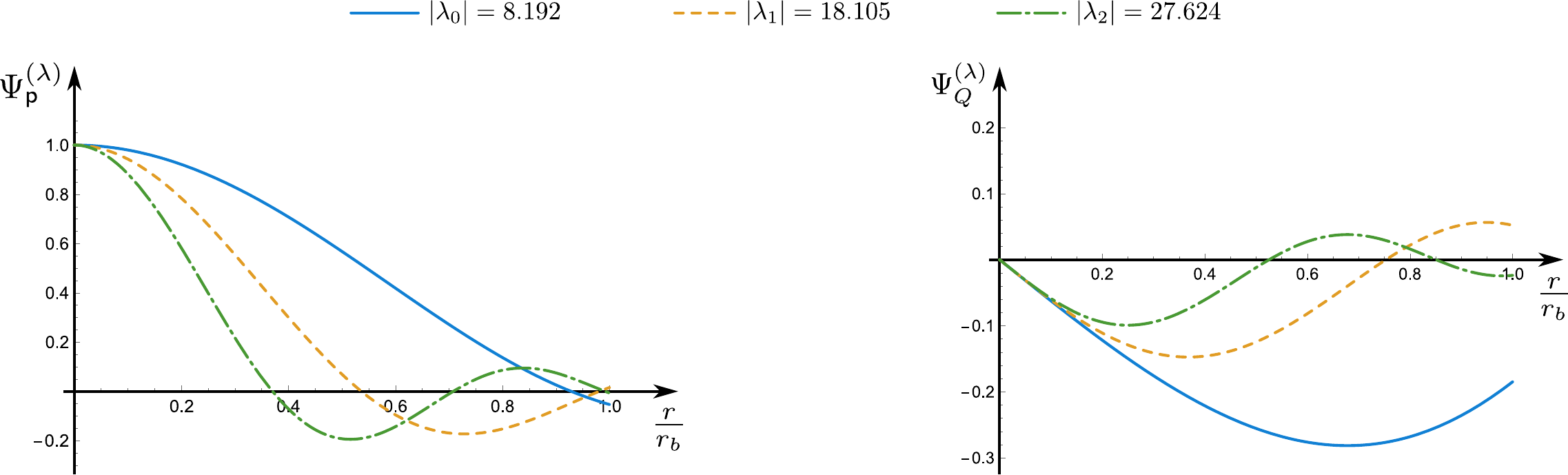}

\caption{\label{fig:Eigenfunctions_PVIIa}Behavior of the radial harmonic
	coefficients of the functions $\mathsf{p}$ and $Q$, related with
	the eigenfrequencies in Table~\ref{table:Fundamental-modes} for the P-V IIa
	spacetime.  It was assumed
	$c_{2}=1$ for all eigenmodes.}
\end{figure}

\begin{figure}
\centering\includegraphics[width=1\textwidth]{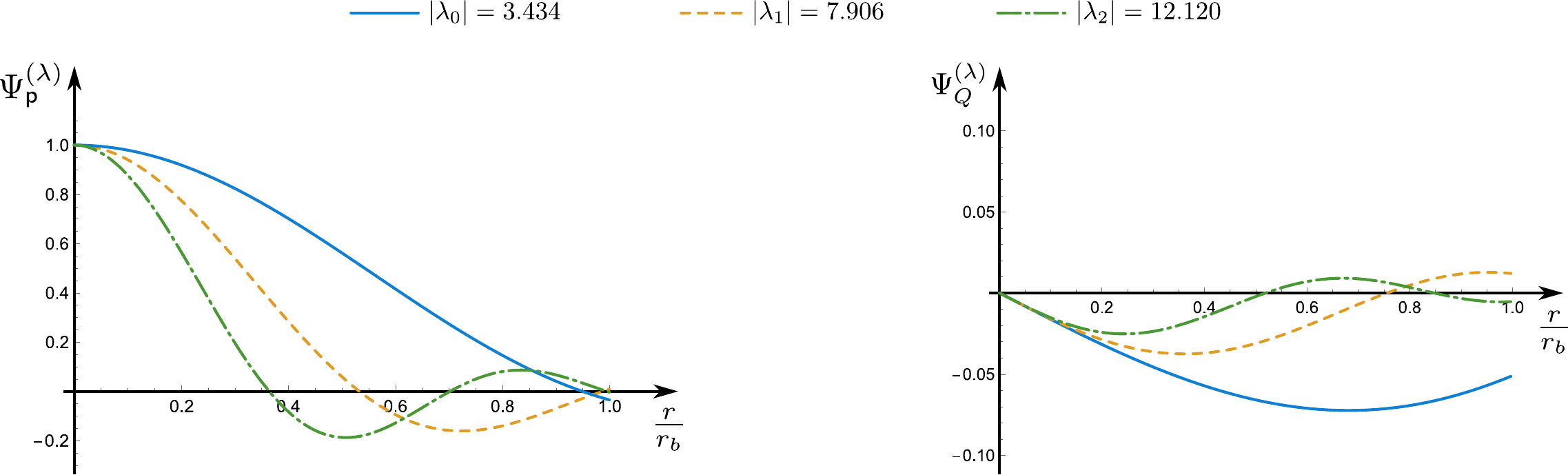}

\caption{\label{fig:Eigenfunctions_TolmanVII}Behavior of the radial harmonic
	coefficients of the functions $\mathsf{p}$ and $Q$, related with
	the eigenfrequencies in Table~\ref{table:Fundamental-modes}
	for the Tolman VII spacetime.  It was assumed
	$c_{2}=1$ for all eigenmodes.}
\end{figure}

In Table~\ref{table:Metric-coefficients}, we present the nontrivial
metric coefficients in the line element~(\ref{eq:general_static_line_element}) for the classical solutions of the EFE mentioned above,  according to the naming conventions
for the solutions of Ref.~\citep{Delgaty_Lake_1998}.
In Table~\ref{table:Fundamental-modes}, we present the absolute values
of the eigenfrequencies associated with the first three eigenmodes
for specific values of the equilibrium background spacetime. Figures~\ref{fig:Eigenfunctions_Buchdahl_I}--\ref{fig:Eigenfunctions_TolmanVII}
show the radial profile of the harmonic coefficients of the functions
$\mathsf{p}$ and $Q$  related with the eigenfrequencies for the
various equilibria in Table~\ref{table:Fundamental-modes}.

In the considered examples, all the eigenfrequencies are real. Hence, for the chosen parameters, all configurations are dynamically stable under adiabatic, radial perturbations.
In Figures~\ref{fig:Eigenfunctions_Buchdahl_I}--\ref{fig:Eigenfunctions_TolmanVII} we find the expected behavior for the radial eigenmodes. For a sufficiently regular background, Eqs.~\eqref{eq:Static_Radial_Adiabatic_pdot_hat} and \eqref{eq:Static_Radial_Adiabatic_boundary_condition_coefficients} can be cast in the form of a Sturm-Liouville eigenvalue problem for $\Psi_Q^{\left(\lambda \right)}$~\cite{Luz_Carloni_2024a}. As a consequence, in particular, the number of roots of the eigenfunctions is associated with the order of the related eigenvalue in the series $\left( \lambda^2_n\right)_{n\in \mathbb{N}}$, as can be inferred from Figures~\ref{fig:Eigenfunctions_Buchdahl_I}--\ref{fig:Eigenfunctions_TolmanVII}.

\begin{table}
\def\arraystretch{1.3}
\begin{tabular}{|c|c|c|c|c|}
\hline 
Spacetime & Parameters & $\left|\lambda_{0}\right|$ & $\left|\lambda_{1}\right|$ & $\left|\lambda_{2}\right|$\tabularnewline
\hline 
\hline 
Buch1 & $\left(A,B,C\right)=\left(1,0.5,1\right)$ & \,\,16.370\,\, & \,\,36.011\,\, & \,\,54.889\,\, \tabularnewline
\hline 
Heint IIa & $\left( a,A,C \right) =\left( 1,1,1.5\right) $  & 4.004 & 10.262 & 15.939\tabularnewline
\hline 
P-V IIa & $\left(A,B,b,c,d\right)=\left(1,3,2,1,1\right)$ & 8.192 & 18.105 & 27.624\tabularnewline
\hline 
\,\,Tolman VII\,\, & $\,\,\left(A,B,C,R\right)=\left(1,1,20,0.54\right)\,\,$ & 3.434 & 7.906 & 12.120\tabularnewline
\hline 
\end{tabular}\caption{\label{table:Fundamental-modes}Absolute values of the eigenfrequencies associated with the first three eigenmodes rounded to
three decimal places, for the equilibria in Table~\ref{table:Metric-coefficients}
assuming selected values of the spacetime parameters.}
\end{table}

The adiabatic, radial perturbation of a background spacetime described
by the Heintzmann IIa solution was previously analyzed in the comoving
frame~\cite{Luz_Carloni_2024b}. We repeated the analysis
here, computing the eigenfunctions and the eigenfrequencies using
the perturbation equations for the static frame to demonstrate the
consistency between both descriptions. As expected, the computed eigenfrequencies
are equal. Indeed, we have compared the values of the eigenfrequencies
associated with the first three eigenmodes for all background spacetimes
in Ref.~\cite{Luz_Carloni_2024b}, using the perturbation
equations for the static frame, Eqs.~(\ref{eq:Static_Radial_Adiabatic_pdot_hat})--(\ref{eq:Static_Radial_Adiabatic_boundary_condition_coefficients}),
finding  complete agreement up to at least 30 significant figures, confirming the
equivalence between the system found for the comoving frame and the
system found for the static frame. We have also compared the results
in Table.~\ref{table:Fundamental-modes} with the predictions of
the systems in Refs.~\citep{Chanmugam_1977,Gondek_1997}. Implementing
a shooting method to numerically integrate the differential equations in those references for the various background spacetimes, the values for the fundamental eigenfrequencies
agree exactly to the considered numerical accuracy, confirming the
equivalence between the new perturbation formalism and the direct
metric-perturbation approach.

\section{Perturbation of neutron stars with a realistic equation of state \label{sec:Perturbation_of_neutron_stars}}

The spacetimes considered in the previous section have the remarkable
property of being real analytic throughout the whole domain of physical
interest. However, this can be different for general background
solutions of the EFE. Indeed, physical solutions are only required
to be twice differentiable. In some cases, even solutions in the weaker
sense can be considered. For such spacetimes, the algorithm presented
in Section~\ref{sec:AdiabaticRadialPerturbations} is not applicable, 
and other methods have to be used, in most cases numerical methods. To illustrate
this approach using the new perturbation equations, we will study
adiabatic, radial perturbations of equilibrium background solutions
with a matter fluid source verifying a ``realistic'' equation of
state suitable to model the interior of cold neutron stars.

Numerous equations of state have been proposed to describe matter
in the high-density regime, considering different effective models
for the interactions between nuclei and various particle species.
Then, using the proposed equations of state, tabulated values for
the matter density, baryon number, and pressure are presented. To use
those values, an interpolation procedure is necessary to generate
a one-parameter barotropic equation of state. This procedure, however,
introduces a source of indeterminacy, such that the eigenfrequencies
may diverge significantly depending on the interpolation scheme, and
direct comparison with the results in the literature is not possible~\citep{Kokkotas_Ruoff_2001}.
Alternatively, significant effort has been made to find analytical representations
of unified equations of state, providing closed models for the matter
fluid within a neutron star~\citep{Haensel_Potekhin_2004,Potekhin_et_al_2013,Suleiman_et_al_2022}.
This is a more sensible approach, also because the various strata
within a neutron star are composed of matter in different regimes.
However, to our knowledge, there has been no comprehensive study of adiabatic
radial perturbations for spacetimes with those types of fluids that
collects the eigenfrequencies for various equilibrium configurations.
As such, we will consider the analytically tractable example of a
fluid characterized by the equation of state given by the Bethe-Johnson
model I within the whole neutron star, since this does not require
the usage of interpolation~\citep{Bethe_Johnson_1974}.

To ease the comparison between our results and those in the literature,
in this section, we will adopt a different unit system to express the
various quantities.

\subsection{The Bethe-Johnson model I and properties of the equilibrium solutions}

Let $E$ represent the energy per neutron measured in MeV,
$n$ the particle density measured in $\text{fm}^{-3}$, $m_{n}$
the rest energy of the neutron in MeV, and $p$ the neutron pressure
measured in MeV per $\text{fm}^{3}$. Then, the Bethe-Johnson model~I
is characterized by the following relations
\begin{equation}
\begin{aligned}E & =236n^{1.54}+m_{n}\,,\\
p & =363.44n^{2.54}\,,
\end{aligned}
\end{equation}
valid for particle densities $0.1\apprle n\apprle3$ $\text{fm}^{-3}$, or
mass densities,
\begin{equation}
1.7\times10^{14}\apprle\rho\apprle1.1\times10^{16}\;\text{g/cm}^{3}\,.
\end{equation}

\begin{figure}
\subfloat[\label{subfig:Mass_density}]{\includegraphics[height=0.19\textheight]{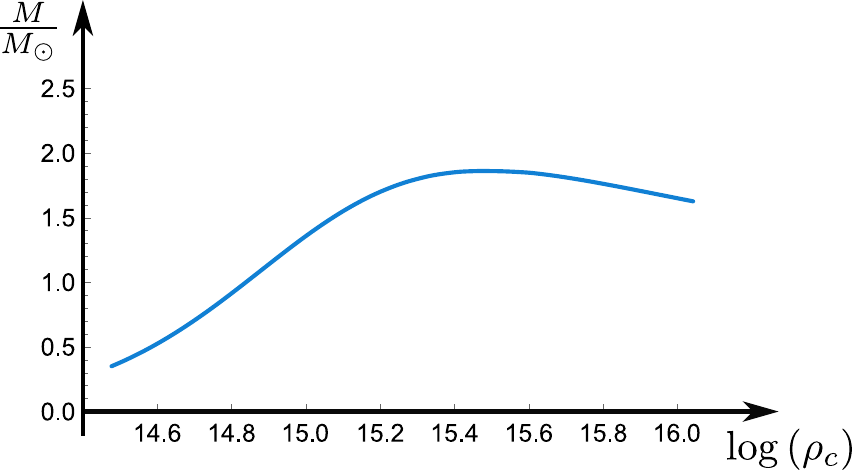}}
\hspace*{\fill}
\subfloat[\label{subfig:Radius_density}]{\includegraphics[height=0.19\textheight]{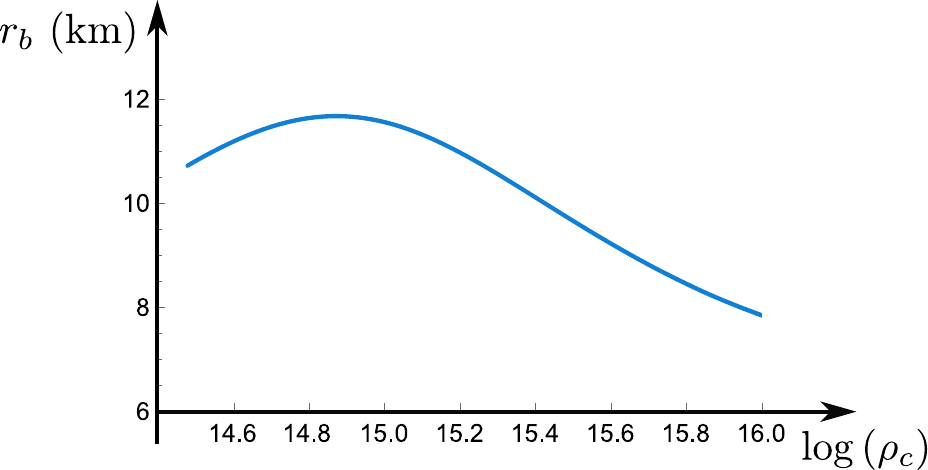}}

\caption{\label{fig:Mass_radius_density}Gravitational mass and radius of the
neutron star as functions of the central matter density for the Bethe-Johnson
model I. The central density $\rho_{c}$ is measured in $\text{g/cm}^{3}$, and $\log$ represents the logarithm base 10. }

\end{figure}

Notice, in particular, that in this section we do not use the geometrized unit system. Therefore, the energy density, $\mu$, and the mass density, $\rho$, have to be explicitly discriminated. Considering a fluid described by this model, providing the central density as initial data,
we can numerically solve the TOV equations to find the equilibrium spacetime.
Since the canonical version of the TOV equations are ill-conditioned and are known to form a numerically stiff system for
certain equations of state, we have verified the results using both
the original version and the reformulation introduced in Ref.~\citep{Lindblom_1998}.
In Fig.~\ref{fig:Mass_radius_density}, we plot the behavior of the
total gravitational mass and the radius of the equilibrium stellar
object for various values of the central matter density. Comparing
the results with those in Ref.~\citep{Malone_Johnson_Bethe_1975},
we see a similar behavior for the total gravitational mass. In contrast,
there is a marked discrepancy between the behavior and the values
of the radius. For instance, we have found the maximum radius to be
around 11.67 km, whereas in Ref.~\citep{Malone_Johnson_Bethe_1975}
the maximum radius is clearly above 12 km. This shift in the values
of the radii is a consequence of the sensitivity of the model to the
values considered for the fundamental constants. In particular, the
radius of the neutron star strongly depends on the value of the rest
energy of the neutron. By considering slightly less accurate values
for $m_{n}$, we confirm that there is an overall increase in the
radius, such that the maximum radius indeed increases above 12
km. On the other hand, we are not able to explain the behavior of
the radius that we see in Fig.~\ref{subfig:Radius_density} for lower
values of the central matter density, where the radius first increases to
a maximum, then monotonically decreases, whereas, in Ref.~\citep{Malone_Johnson_Bethe_1975},
the radius monotonically decreases for all considered values of the
central matter density. We speculate that this is a consequence of
the number of points considered in that region in Ref~\citep{Malone_Johnson_Bethe_1975}.

\subsection{Oscillation eigenfrequencies and $e$-folding time for adiabatic
radial perturbations}

Defining the background spacetime, we can numerically integrate the
perturbation equations~(\ref{eq:Static_Radial_Adiabatic_pdot_hat})
and (\ref{eq:Static_Radial_Adiabatic_Q_hat}) imposing the boundary
conditions \ref{enu:general_boundary_condition_1} and \ref{enu:general_boundary_condition_2}
to find the values of the eigenfrequencies.

In Table.~\ref{table:Bethe_Johnson_fundamental-modes}, we present
the values of the oscillation frequencies or the $e$-folding time 
associated with the first three
eigenmodes for various values of the central mass density. Comparing
with the results of Ref.~\citep{Kokkotas_Ruoff_2001}, there is a
disagreement between the values of the various quantities. In particular,
we find that, for the same values of the central density, the neutron
star is more compact than in Ref.~\citep{Kokkotas_Ruoff_2001} (see fourth column of Table~\ref{table:Bethe_Johnson_fundamental-modes}). However, the difference between
the results is explained by the accuracy 
of the values of the fundamental constants. Nevertheless, our conclusions regarding
the stability are in perfect agreement with those of Ref.~\citep{Kokkotas_Ruoff_2001}.
To attest the consistency of our results, we confirmed that a zero
value for the eigenfrequency of the fundamental mode occurs exactly
for the value of the central matter density that leads to the maximum
gravitational mass. This is the expected result since we have considered
that a fluid with the same equation of state permeates both the background and the perturbed spacetimes. The consistency of our
approach was further confirmed by comparing the predictions of the
new perturbation scheme with those of metric-based perturbation theory,
using the systems introduced in Refs.~\citep{Chanmugam_1977,Gondek_1997}.

\begin{table}
\def\arraystretch{1.3}
\begin{tabular}{|c|c|c|c|c|c|c|}
\hline 
~\raisebox{11pt}{\phantom{.}}$\rho_{c}~\left(10^{15}\text{g/cm}^{3}\right)$\raisebox{-7pt}{\phantom{.}}~ & ~$r_{b}$ (km)~ & ~~~$\frac{M}{M_{\odot}}$~~~ & ~~~$\frac{GM}{c^{2}r_b}$~~~ & ~$f_{1}$ $\left(\text{kHz}\right)$~ & ~$f_{2}$ $\left(\text{kHz}\right)$~ & ~$f_{3}$ $\left(\text{kHz}\right)$~\tabularnewline
\hline 
\hline
3.10 & 9.692 & 1.864 & 0.284 & ~1.066{*} & 18.658 & 28.835\tabularnewline
3.05 & 9.724 & 1.865 & 0.283 & 0.647 & 18.678 & 28.536\tabularnewline
2.80 & 9.891 & 1.862 & 0.278 & 2.366 & 17.550 & 27.008\tabularnewline
2.50 & 10.115 & 1.851 & 0.270 & 3.270 & 16.378 & 25.093\tabularnewline
2.00 & 10.545 & 1.801 & 0.252 & 3.998 & 14.266 & 21.678\tabularnewline
1.50 & 11.058 & 1.669 & 0.223 & 4.147 & 11.916 & 17.940\tabularnewline
1.00 & 11.554 & 1.358 & 0.174 & 3.800 & 9.271 & 13.809\tabularnewline
\hline 
\end{tabular}\caption{\label{table:Bethe_Johnson_fundamental-modes}Oscillation frequencies
or $e$-folding time of adiabatic, radial perturbations of equilibrium
spacetimes with a matter fluid verifying the Bethe-Johnson model I equation
of state. From left to right, we list the central mass density,
$\rho_{c}$, radius of the neutron star, $r_{b}$, gravitational mass,
$M$, the compactness parameter, $\frac{GM}{c^{2}r_b}$, 
and the first three oscillation frequencies, $f_{i}=\lambda_{i}/\left(2\pi\right)$.
In the second row, the entry with an asterisk indicates the $e$-folding
time in $ms$ for the fundamental mode.}
\end{table}

\section{Conclusion\label{sec:Conclusions}}

We have studied adiabatic, radial perturbations of static, self-gravitating
perfect fluids within the theory of general relativity employing the
new perturbative formalism introduced in Ref.~\cite{Luz_Carloni_2024a},
by considering the point of view of radially static observers. For
these observers, the fluid is no longer perceived as being perfect, and momentum transfer terms have to be included in the stress-energy
tensor. We discussed in detail the thermodynamic description of the
matter fluid in the static frame, showing explicitly that at linear
order the perturbations are still adiabatic. Nonetheless,
the covariance of the formalism allows us to deduce easily
that a radially static observer will measure an apparent entropy flux between the volume elements of	the fluid.
	
This result has to be considered carefully. 
	
In general, in a non-equilibrium state, it is not necessarily possible to describe thermodynamical systems with the state variables considered for systems in equilibrium. To this end, we have considered the local-thermodynamic equilibrium ansatz.
Notice, however, that this ansatz is only reasonable if the oscillation evolution timescale is much bigger than a characteristic local relaxation time of the fluid sources.
This leads to a limitation in the applicability of thermodynamical arguments for modes associated with eigenfrequencies that are not small enough. Such a caveat should always be considered with care, as nontrivial thermodynamics, in general, will influence the dynamical evolution of perturbations. 
Beyond this limit, without a complete theory of non-equilibrium thermodynamics, it is impossible to predict 
the thermodynamic evolution of the perturbed matter fluid.
This discussion shows how important it would be to measure the oscillation modes of compact stellar objects directly. These data would offer a window into the behavior of matter in strong relativistic regimes and guide the development of such a theory.
If we accept that the static observer will measure a true entropy flow, we can deduce that it will be possible to define an apparent temperature for the fluid. Unfortunately, the analysis of the entropy flow does not bring any information on the higher-order corrections to the temperature.  Our inability to determine these corrections ultimately stems from the choice of the state variables. In particular, the fact that the local temperature appears as an integration factor multiplying a first-order variation of a state variable implies that higher-order temperature corrections cannot appear in any perturbative analysis employing the local-thermodynamical equilibrium ansatz.
In other words, the very assumption that a fluid out of thermodynamic equilibrium depends locally on the same state variables verifying the same relations as if the system were in a state of thermodynamic equilibrium effectively prevents us from obtaining information on the higher-order corrections to the temperature and even their very meaning.

Although the thermodynamic description of the fluid in the static frame is more complex
than the one in the fluids' local rest frame, we have shown that its dynamical description
is greatly simplified.
Indeed, analytic solutions were developed for
a general class of equilibrium spacetimes, leading to a significantly
more efficient algorithm to find power series solutions when compared
with the results found from the point of view of comoving observers~\cite{Luz_Carloni_2024b}.
To illustrate the new system of equations, we have analyzed adiabatic,
radial perturbations of selected non-trivial exact solutions of the
theory of general relativity for the equilibrium spacetime, and computed
the first eigenfrequencies and the corresponding eigenfunctions. We
have discussed how the results compare with the predictions of metric-based
perturbation formalisms for the same spacetimes, finding complete
agreement to the considered accuracy.

Lastly, we have analyzed adiabatic, radial perturbations
of cold neutron stars composed of a perfect fluid characterized by the Bethe-Johnson
model I equation of state. In this case, the background spacetime
is not real analytic, and the exact power-series solutions are not
applicable. Nonetheless, the example illustrates that the system
is well-conditioned and suitable for numerical methods. Indeed, we
have studied the eigenfrequencies of the perturbed fluid, showing
their consistency with those found using metric-based perturbation
frameworks in the literature.

\appendix

\section{\label{Appendix:sec_1p1p2_decomposition}The 1+1+2 decomposition}

\subsection{Projectors and the Levi-Civita volume form}

Assume the existence of an affinely parameterized congruence of timelike curves with normalized tangent vector
field $u$ in some open neighborhood of a Lorentzian manifold of dimension 4, $\left(\mathcal{M},g\right)$, where $g$ is a metric tensor with components 
$g_{\alpha\beta}$. We can use the vector field $u$ to locally foliate the manifold in 3-surfaces, such that any tensor quantity can be pointwise decomposed in its projection along the direction of $u$ and onto the tangent and cotangent space of a 3-surface $V$. Such decomposition is called 1+3 formalism and relies on a projector tensor $h$. Using the metric tensor $g$ and the vector field $u$, we can naturally define $h$ such that
\begin{equation}
h_{\alpha\beta}=g_{\alpha\beta}+u_{\alpha}u_{\beta}\,,
\label{Def_eq:projector_h_definition}
\end{equation}
where $u_\alpha$ represent the components of the 1-form associated with $u$, with the properties
\begin{equation}
\begin{aligned}h_{\alpha\beta} & =h_{\beta\alpha}\,, &  &  & h_{\alpha\beta}h^{\beta\gamma} & =h_{\alpha}{}^{\gamma}\,,\\
h_{\alpha\beta}u^{\alpha} & =0\,, &  &  & h_{\alpha}{}^{\alpha} & =3\,.
\end{aligned}
\end{equation}

In addition to the timelike congruence, we assume the existence of another congruence of spacelike curves with normalized tangent vector field $e$ such
that, in analogy with the 1+3 decomposition, any tensor quantity defined in the submanifold $V$ can be decomposed along $e$ and the 2-surfaces $W$, orthogonal to
both $u$ and $e$. This defines the 1+1+2 covariant formalism. This formalism relies on the existence of a projector, $N$, onto the cotangent space of $W$. Considering the metric tensor and the vector fields $u$ and $e$, we define
\begin{equation}
N_{\alpha\beta}=h_{\alpha\beta}-e_{\alpha}e_{\beta}\,,
\label{Def_eq:projector_N_definition}
\end{equation}
where $e_\alpha$ represents the components of the 1-form associated with $e$. This operator has the following properties
\begin{equation}
\begin{aligned}N_{\alpha\beta} & =N_{\beta\alpha}\,, &  &  & N_{\alpha\beta}N^{\beta\gamma} & =N_{\alpha}{}^{\gamma}\,,\\
N_{\alpha\beta}u^{\alpha} & =N_{\alpha\beta}e^{\alpha}=0\,, &  &  & N_{\alpha}{}^{\alpha} & =2\,.
\end{aligned}
\label{Def_eq:projector_N_properties}
\end{equation}

To complete the representation of the geometry of $\left(\mathcal{M},g\right)$ we introduce the totally skew-symmetric
tensors
\begin{equation}
\begin{aligned}\varepsilon_{\alpha\beta\gamma} & =\varepsilon_{\alpha\beta\gamma\sigma}u^{\sigma}\,,\\
\varepsilon_{\alpha\beta} & =\varepsilon_{\alpha\beta\gamma}e^{\gamma}\,.
\end{aligned}
\label{Def_eq:volume_forms}
\end{equation}
 where $\varepsilon_{\alpha\beta\gamma\sigma}$ is the covariant Levi-Civita tensor. 
Lastly, we will indicate the symmetric and
anti-symmetric part of a 2-tensor $\chi$ using parentheses and brackets, such
that
\begin{equation}
\begin{aligned}\chi_{\left(\alpha\beta\right)} & =\frac{1}{2}\left(\chi_{\alpha\beta}+\chi_{\beta\alpha}\right)\,, &  &  & \chi_{\left[\alpha\beta\right]} & =\frac{1}{2}\left(\chi_{\alpha\beta}-\chi_{\beta\alpha}\right)\,.\end{aligned}
\end{equation}

\subsection{\label{Appendix:subsec_Covariant_derivatives_decomposition}Covariant
derivatives of $u$ and $e$}

The 1+1+2 decomposition can be applied to the covariant derivatives of the vector fields $u$ and $e$ and define the kinematical quantities associated with the respective congruences.
To lighten the notation, given a tensor quantity $\chi$, we will use
\begin{equation}
	\begin{aligned}
		\dot{\chi} & :=u^{\mu} \nabla_\mu \chi\,, &  &  & \widehat{\chi} & :=e^{\mu} \nabla_\mu \chi\,,
	\end{aligned}
\end{equation}
for the directional derivatives along the integral curves of $u$ and the integral curves of $e$, respectively.

Now, the covariant derivative of the vector field
$u$ can be decomposed as
\begin{equation}
\nabla_{\alpha}u_{\beta}=-u_{\alpha}\left(\mathcal{A}e_{\beta}+\mathcal{A}_{\beta}\right)+\frac{1}{3}h_{\alpha\beta}\theta+\sigma_{\alpha\beta}+\omega_{\alpha\beta}\,,
\end{equation}
where
\begin{equation}
\begin{aligned}\mathcal{A} & =-u_{\mu}u^{\nu}\nabla_{\nu}e^{\mu}\,, &  &  & \mathcal{A}_{\alpha} & =N_{\alpha\mu}\dot{u}^{\mu}\,,\end{aligned}
\end{equation}
and $\theta$ represents the expansion scalar, $\sigma_{\alpha\beta}$ the shear tensor, and $\omega_{\alpha\beta}$ the vorticity tensor,
such that
\begin{equation}
\begin{aligned}\theta & =h^{\mu\nu}\nabla_{\mu}u_{\nu}\,,\\
\sigma_{\alpha\beta} & =\left(\frac{h_{\alpha}{}^{\mu}h_{\beta}{}^{\nu}+h_{\alpha}{}^{\nu}h_{\beta}{}^{\mu}}{2}-\frac{1}{3}h_{\alpha\beta}h^{\mu\nu}\right)\nabla_{\mu}u_{\nu}=\Sigma_{\alpha\beta}+2\Sigma_{(\alpha}e_{\beta)}+\Sigma\left(e_{\alpha}e_{\beta}-\frac{1}{2}N_{\alpha\beta}\right)\,,\\
\omega_{\alpha\beta} & =\frac{1}{2}\left(h_{\alpha}{}^{\mu}h_{\beta}{}^{\nu}-h_{\alpha}{}^{\nu}h_{\beta}{}^{\mu}\right)\nabla_{\mu}u_{\nu}=\varepsilon_{\alpha\beta\mu}\left(\Omega e^{\mu}+\Omega^{\mu}\right)\,,
\end{aligned}
\end{equation}
with
\begin{equation}
\begin{aligned}\Sigma_{\alpha\beta} & =\left(\frac{N_{\alpha}{}^{\mu}N_{\beta}{}^{\nu}+N_{\alpha}{}^{\nu}N_{\beta}{}^{\mu}}{2}-\frac{1}{2}N_{\alpha\beta}N^{\mu\nu}\right)\sigma_{\mu\nu}\,, &  &  & \Sigma_{\alpha} & =N_{\alpha}{}^{\mu}e^{\nu}\sigma_{\mu\nu}\,, &  &  & \Sigma & =e^{\mu}e^{\nu}\sigma_{\mu\nu}\,,\end{aligned}
\end{equation}
and
\begin{equation}
\begin{aligned}\Omega^{\alpha} & =\frac{1}{2}N_{\gamma}{}^{\alpha}\varepsilon^{\mu\nu\gamma}\nabla_{\mu}u_{\nu}\,, &  &  & \Omega & =\frac{1}{2}\varepsilon^{\mu\nu}\nabla_{\mu}u_{\nu}\,.\end{aligned}
\end{equation}
From their definitions, we see that vector and 2-tensor quantities
characterize the behavior of the kinematical quantities on the 2-surfaces
$W$, and the scalars characterize their behavior along $u$ or
$e$.

Similarly, the covariant derivative of the vector field $e$ can be written as the following sum
\begin{equation}
\nabla_{\alpha}e_{\beta}=\frac{1}{2}N_{\alpha\beta}\phi+\zeta_{\alpha\beta}+\varepsilon_{\alpha\beta}\xi+e_{\alpha}a_{\beta}-u_{\alpha}\alpha_{\beta}-\mathcal{A}u_{\alpha}u_{\beta}+\left(\frac{1}{3}\theta+\Sigma\right)e_{\alpha}u_{\beta}+\left(\Sigma_{\alpha}-\varepsilon_{\alpha\mu}\Omega^{\mu}\right)u_{\beta}\,,\label{Def_eq:Cov_dev_vector_e-1}
\end{equation}
where
\begin{equation}
\begin{aligned}\phi & =N^{\mu\nu}\nabla_{\mu}e_{\nu}\,, &  &  & \zeta_{\alpha\beta} & =\left(\frac{N_{\alpha}{}^{\mu}N_{\beta}{}^{\nu}+N_{\alpha}{}^{\nu}N_{\beta}{}^{\mu}}{2}-\frac{1}{2}N_{\alpha\beta}N^{\mu\nu}\right)\nabla_{\mu}e_{\nu}\,, &  &  & \xi & =\frac{1}{2}\varepsilon^{\mu\nu}\nabla_{\mu}e_{\nu}\,,\end{aligned}
\end{equation}
are the kinematical quantities of the congruence associated with the vector field $e$ on $W$, namely, $\phi$ is the expansion scalar, $\zeta_{\alpha \beta}$ the shear tensor, and
the $\xi$ twist, and
\begin{equation}
\begin{aligned}a_{\alpha} & =e^{\mu}h_{\alpha}{}^{\nu}\nabla_{\mu}e_{\nu}\,, &  &  & \alpha_{\alpha} & =u^{\mu}h_{\alpha}{}^{\nu}\nabla_{\mu}e_{\nu}\,.\end{aligned}
\end{equation}

\subsection{Weyl tensor}

In covariant approaches, the conformal structure of the spacetime plays a fundamental role. For this reason, the Weyl 
tensor, with components $C_{\alpha\beta\gamma\delta}$, which describes the tidal forces and the properties of gravitational waves, needs to be expressed in terms of 1+1+2 variables.
As is well known, the Weyl tensor can be defined as the trace-free part of the Riemann curvature tensor, $R_{\alpha\beta\gamma\delta}$, such that, in four spacetime dimensions, we have
\begin{equation}
R_{\alpha\beta\gamma\delta}=C_{\alpha\beta\gamma\delta}+R_{\alpha\left[\gamma\right.}g_{\left.\delta\right]\beta}-R_{\beta\left[\gamma\right.}g_{\left.\delta\right]\alpha}-\frac{1}{3}R\,g_{\alpha\left[\gamma\right.}g_{\left.\delta\right]\beta}\,.
\label{Def_eq:Weyl_tensor_definition}
\end{equation}
Remarkably,
we need only two 2-tensors to characterize the Weyl 4-tensor in general relativity completely:
\begin{align}
E_{\alpha\beta} & =C_{\alpha\mu\beta\nu}u^{\mu}u^{\nu}\,,\label{Def_eq:Weyl_tensor_electric}\\
H_{\alpha\beta} & =\frac{1}{2}\varepsilon_{\alpha}{}^{\mu\nu}C_{\mu\nu\beta\delta}u^{\delta}\,,
\end{align}
where $E$ is called ``electric'' part of the Weyl tensor, while  $H$ is called ``magnetic'' part of the Weyl tensor. They are both symmetric and traceless tensors, such that
\begin{equation}
C_{\alpha\beta\gamma\delta}=-\varepsilon_{\alpha\beta\mu}\varepsilon_{\gamma\delta\nu}E^{\nu\mu}-2u_{\alpha}E_{\beta\left[\gamma\right.}u_{\left.\delta\right]}+2u_{\beta}E_{\alpha\left[\gamma\right.}u_{\left.\delta\right]}-2\varepsilon_{\alpha\beta\mu}H^{\mu}{}_{\left[\gamma\right.}u_{\left.\delta\right]}-2\varepsilon_{\mu\gamma\delta}H^{\mu}{}_{\left[\alpha\right.}u_{\left.\beta\right]}\,.
\label{Def_eq:Weyl_tensor_1+3_decomposition}
\end{equation}
This represents the famous 1+3 decomposition of the Weyl tensor.
In the 1+1+2 spacetime decomposition formalism, the electric and magnetic parts of the Weyl tensor can be further decomposed as 
\begin{equation}
\begin{aligned}E_{\alpha\beta} & =\mathcal{E}\left(e_{\alpha}e_{\beta}-\frac{1}{2}N_{\alpha\beta}\right)+\mathcal{E}_{\alpha}e_{\beta}+e_{\alpha}\mathcal{E}_{\beta}+\mathcal{E}_{\alpha\beta}\,,\\
H_{\alpha\beta} & =\mathcal{H}\left(e_{\alpha}e_{\beta}-\frac{1}{2}N_{\alpha\beta}\right)+\mathcal{H}_{\alpha}e_{\beta}+e_{\alpha}\mathcal{H}_{\beta}+\mathcal{H}_{\alpha\beta}\,,
\end{aligned}
\end{equation}
where
\begin{equation}
\begin{aligned}\mathcal{E} & =E_{\mu\nu}e^{\mu}e^{\nu}=-N^{\mu\nu}E_{\mu\nu}\,, &  &  & \mathcal{H} & =e^{\mu}e^{\nu}H_{\mu\nu}=-N^{\mu\nu}H_{\mu\nu}\,,\\
\mathcal{E}_{\alpha} & =N_{\alpha}{}^{\mu}e^{\nu}E_{\mu\nu}=e^{\mu}N_{\alpha}{}^{\nu}E_{\mu\nu}\,, &  &  & \mathcal{H}_{\alpha} & =N_{\alpha}{}^{\mu}e^{\nu}H_{\mu\nu}=e^{\mu}N_{\alpha}{}^{\nu}H_{\mu\nu}\,,\\
\mathcal{E}_{\alpha\beta} & =E_{\left\{ \alpha\beta\right\} }\,, &  &  & \mathcal{H}_{\alpha\beta} & =H_{\left\{ \alpha\beta\right\} }\,.
\end{aligned}
\label{Def_eq:Weyl_tensor_components_definition}
\end{equation}

\subsection{Stress-energy tensor}

Lastly, to complete the 1+1+2 gravitational equations, we also need a decomposition of the stress-energy tensor, with components $T_{\alpha\beta}$. Using the projector operators~\eqref{Def_eq:projector_h_definition} and \eqref{Def_eq:projector_N_definition} yields
\begin{equation}
T_{\alpha\beta}=\mu\,u_{\alpha}u_{\beta}+\left(p+\Pi\right)e_{\alpha}e_{\beta}+\left(p-\frac{1}{2}\Pi\right)N_{\alpha\beta}+2Qe_{\left(\alpha\right.}u_{\left.\beta\right)}+2Q_{\left(\alpha\right.}u_{\left.\beta\right)}+2\Pi_{\left(\alpha\right.}e_{\left.\beta\right)}+\Pi_{\alpha\beta}\,,\label{Def_eq:Stress-energy_tensor_decomposition}
\end{equation}
with 
\begin{equation}
\begin{aligned}\mu & =u^{\mu}u^{\nu}T_{\mu\nu}\,, &  &  & Q_{\alpha} & =-N_{\alpha}{}^{\mu}u^{\nu}T_{\mu\nu}\,,\\
p & =\frac{1}{3}\left(e^{\mu}e^{\nu}+N^{\mu\nu}\right)T_{\mu\nu}\,, &  &  & \Pi_{\alpha} & =N_{\alpha}{}^{\mu}e^{\nu}T_{\mu\nu}\,,\\
\Pi & =\frac{1}{3}\left(2e^{\mu}e^{\nu}-N^{\mu\nu}\right)T_{\mu\nu}\,, &  &  & \Pi_{\alpha\beta} & =\left(\frac{N_{\alpha}{}^{\mu}N_{\beta}{}^{\nu}+N_{\alpha}{}^{\nu}N_{\beta}{}^{\mu}}{2}-\frac{1}{2}N_{\alpha\beta}N^{\mu\nu}\right)T_{\mu\nu}\,.\\
Q & =-e^{\mu}u^{\nu}T_{\mu\nu}\,,
\end{aligned}
\label{Def_eq:Energy_momentum_tensor_decomposition_quantities}
\end{equation}
For an observer with 4-velocity $u$,
$\mu$ represents the perceived mass-energy density of the fluid,
$p$ the isotropic pressure, $Q$ characterizes the energy-momentum flow along $e$, $Q_{\alpha}$ is the energy-momentum flux in $W$, and $\Pi$,
$\Pi_{\alpha}$ and $\Pi_{\alpha\beta}$ characterize the anisotropic
pressure of the matter fluid.

\section{\label{Appendix:sec_isotropic_frame_transformations}Change in the stress-energy
tensor under isotropic frame transformations}

In this appendix, we will define an isotropic frame transformation
associated with two dyads, and discuss how this type of transformation
changes the stress-energy tensor used to describe the fluid in each
frame.

\subsection{Isotropic frame transformations and projectors}

Let a local frame be partially defined by a dyad $\left(u,e\right)$ composed, respectively, by a timelike
and a spacelike vector field in a spacetime, such that $u^{\alpha}u_{\alpha}=-1$
and $e^{\alpha}e_{\alpha}=+1$. Then, let another frame be partially defined by another dyad $\left(\bar{u},\bar{e}\right)$ also formed, respectively, by a timelike
and a spacelike vector field, such that $\bar{u}^{\alpha}\bar{u}_{\alpha}=-1$
and $\bar{e}^{\,\alpha}\bar{e}_{\,\alpha}=+1$. As explained in detail in Ref.~\cite{Luz_Carloni_2024a}, a general isotropic frame transformation can be represented by the following relations
\begin{equation}
\begin{aligned}\bar{u}^{\alpha} & =u^{\alpha}\cosh\beta+e^{\alpha}\sinh\beta\,,\\
\bar{e}^{\,\alpha} & =u^{\alpha}\sinh\beta+e^{\alpha}\cosh\beta\,,
\end{aligned}
\label{eq:Frame_transformations_ue_Bar_to_noBar}
\end{equation}
or, equivalently,
\begin{equation}
\begin{aligned}u^{\alpha} & =\bar{u}^{\alpha}\cosh\beta-\bar{e}^{\,\alpha}\sinh\beta\,,\\
e^{\alpha} & =-\bar{u}^{\alpha}\sinh\beta+\bar{e}^{\,\alpha}\cosh\beta\,,
\end{aligned}
\label{eq:Frame_transformations_ue_noBar_to_Bar}
\end{equation}
where $\beta$ is called the tilting angle between the frames.

Using these relations, we can find the transformation of the projector
tensor $h$, defined in Eq.~\eqref{Def_eq:projector_h_definition},
\begin{equation}
\begin{aligned}\bar{h}_{\alpha\beta} & =h_{\alpha\beta}+\left(u_{\alpha}u_{\beta}+e_{\alpha}e_{\beta}\right)\sinh^{2}\beta+\frac{1}{2}\left(u_{\alpha}e_{\beta}+e_{\alpha}u_{\beta}\right)\sinh\left(2\beta\right)\,,\\
h_{\alpha\beta} & =\bar{h}_{\alpha\beta}+\left(\bar{u}_{\alpha}\bar{u}_{\beta}+\bar{e}_{\alpha}\bar{e}_{\beta}\right)\sinh^{2}\beta-\frac{1}{2}\left(\bar{u}_{\alpha}\bar{e}_{\beta}+\bar{e}_{\alpha}\bar{u}_{\beta}\right)\sinh\left(2\beta\right)\,,
\end{aligned}
\label{eq:Frame_transformations_projector}
\end{equation}
with the following properties
\begin{equation}
\begin{aligned}\bar{h}_{\alpha\beta}\bar{u}^{\alpha} & =0\,,\\
\bar{h}_{\alpha\beta}\bar{e}^{\,\alpha} & =\bar{e}_{\beta}\,.
\end{aligned}
\end{equation}
Moreover, we find $\bar{N}_{\alpha\beta}=N_{\alpha\beta}$, such that
$\bar{N}_{\alpha\beta}\bar{u}^{\beta}=0$ and $\bar{N}_{\alpha\beta}\bar{e}^{\,\beta}=0$.

\subsection{Tilting angle, the stress-energy tensor, and the expansion scalar}

We are now interested in relating the tilting angle $\beta$ of an
isotropic frame transformation with the thermodynamic variables of
the fluid measured in the resultant frame.

Consider a metric stress-energy tensor of an isotropic fluid decomposed
accordingly with Eq.~(\ref{Def_eq:Stress-energy_tensor_decomposition}),
that is, the vector and tensor components $Q_{\alpha}=\Pi_{\alpha}=\Pi_{\alpha\beta}=0$.
Under the transformation~(\ref{eq:Frame_transformations_ue_Bar_to_noBar})
we have~~~~
\begin{equation}
\begin{aligned}\bar{\mu} & =\mu-Q\sinh\left(2\beta\right)+\left(\mu+p+\Pi\right)\sinh^{2}\beta\,,\\
\bar{p} & =p-\frac{1}{3}Q\sinh\left(2\beta\right)+\frac{1}{3}\left(\mu+p+\Pi\right)\sinh^{2}\beta\,,\\
\bar{Q} & =Q\cosh\left(2\beta\right)-\frac{1}{2}\left(\mu+p+\Pi\right)\sinh\left(2\beta\right)\,,\\
\bar{\Pi} & =\Pi\left(1+\frac{2}{3}\sinh^{2}\beta\right)-\frac{2}{3}Q\sinh\left(2\beta\right)+\frac{2}{3}\left(\mu+p\right)\sinh^{2}\beta\,.
\end{aligned}
\label{Isotropic_frame_transformations_eq:mu_p_Q_frame_transformations_general}
\end{equation}
Here, an overline characterizes variables measured in the tilted frame
$\left(\bar{u},\bar{e}\right)$.

Now, assuming that the stress-energy tensor $T_{\alpha\beta}$, associated
with the dyad $\left(u,e\right)$, is such that $Q=0$ and $\Pi=0$,
that is, in the frame associated with that dyad, the fluid can be described
by a perfect fluid model, we find that in the resultant barred frame
the fluid is characterized by
\begin{equation}
T_{\alpha\beta}=\bar{\mu}\,\bar{u}_{\alpha}\bar{u}_{\beta}+\left(\bar{p}+\bar{\Pi}\right)\bar{e}_{\alpha}\bar{e}_{\beta}+\left(\bar{p}-\frac{1}{2}\bar{\Pi}\right)\bar{N}_{\alpha\beta}+2\bar{Q}\bar{e}_{\left(\alpha\right.}\bar{u}_{\left.\beta\right)}\,,
\label{Isotropic_frame_transformations_eq:stress_energy_tensor_transformed_perfect_fluid}
\end{equation}
where
\begin{equation}
\begin{aligned}\bar{\mu} & =\mu+\left(\mu+p\right)\sinh^{2}\beta\,,\\
\bar{p} & =p+\frac{1}{3}\left(\mu+p\right)\sinh^{2}\beta\,,\\
\bar{Q} & =-\frac{1}{2}\left(\mu+p\right)\sinh\left(2\beta\right)\,,\\
\bar{\Pi} & =\frac{2}{3}\left(\mu+p\right)\sinh^{2}\beta\,.
\end{aligned}
\label{Isotropic_frame_transformations_eq:mu_p_Q_frame_transformations_perfect}
\end{equation}

For the discussion in the main body of the text, it is also useful
to find the transformation for the expansion scalars measured by the
fiducial curve of each frame: the barred and non-barred frames. Given
the frame transformations~(\ref{eq:Frame_transformations_ue_Bar_to_noBar}),
(\ref{eq:Frame_transformations_ue_noBar_to_Bar}) and (\ref{eq:Frame_transformations_projector})
we find the following relations
\begin{equation}
\bar{\theta}=\theta\cosh\beta+\left(\mathcal{A}+\phi\right)\sinh\beta+u^{\mu}\nabla_{\mu}\left(\cosh\beta\right)+e^{\mu}\nabla_{\mu}\left(\sinh\beta\right)\,.\label{eq:Frame_transformations_thetaBar_theta_general}
\end{equation}

Now, assuming that the fluid that permeates the spacetime can be considered
as a perturbation of a perfect fluid with energy density and pressure
$\mu_{0}$ and $p_{0}$, respectively. Using Eq.~(\ref{Isotropic_frame_transformations_eq:mu_p_Q_frame_transformations_perfect}), we can determine a relation between the tilting angle $\beta$ and
$\bar{Q}$, such that, to linear order of perturbation theory, we find
(cf. Ref.~\cite{Luz_Carloni_2024a}):
\begin{equation}
\beta=-\frac{\bar{Q}}{\mu_{0}+p_{0}}\,.\label{eq:Frame_transformations_beta_heatFlow_linear_order}
\end{equation}
Moreover, the stress-energy tensor, at linear perturbative order, is given simply by
\begin{equation}
		T_{\alpha\beta}=\bar{\mu}\,\bar{u}_{\alpha}\bar{u}_{\beta}+
		\bar{p}\,\left( \bar{e}_{\alpha}\bar{e}_{\beta} + \bar{N}_{\alpha\beta} \right)+
		2\bar{Q}\bar{e}_{\left(\alpha\right.}\bar{u}_{\left.\beta\right)}\,.
		\label{Isotropic_frame_transformations_eq:stress_energy_tensor_transformed_perfect_fluid_simplified}
\end{equation}

\end{document}